\documentclass[aps,pra,reprint,twocolumn,superscriptaddress,longbibliography,nofootinbib,secnumarabic]{revtex4-2}

\usepackage[T1]{fontenc}
\usepackage[utf8]{inputenc}
\usepackage[english]{babel}   
\usepackage{lmodern}
\usepackage{amsmath,amssymb,amsthm,bm,bbm}
\usepackage{siunitx}
\usepackage{xspace}
\usepackage{graphicx}
\usepackage{hyperref}
\usepackage{braket}
\usepackage{physics}        
\usepackage{enumitem}      
\usepackage{times}
\usepackage{listings}
\usepackage{mathtools}
\usepackage[dvipsnames]{xcolor}
\usepackage[normalem]{ulem}
\usepackage{tabularx}
\usepackage{array}
\usepackage{threeparttable}
\usepackage{placeins}
\usepackage{orcidlink}
\usepackage{tcolorbox}

\hypersetup{colorlinks=true,linkcolor=blue,citecolor=blue,urlcolor=blue}
\usepackage{etoolbox}

\definecolor{redcom}{rgb}{0,0,0}
\definecolor{querycol}{rgb}{0,0,0}

\newcommand{\rmedit}[1]{{\color{black}#1}}

\definecolor{purplerep}{rgb}{0,0,0}

\definecolor{green}{rgb}{0,0,0}

\newcommand{\tps}{\tau_{\mathrm{ps}}}
\newcommand{\tpair}{\tau_{\mathrm{pair}}}
\newcommand{\tbn}{\tau_{\mathrm{bn}}}
\newcommand{\Rp}{R_{\mathrm p}}
\newcommand{\Rtrue}{R_{\mathrm{twinp}}}
\newcommand{\Racc}{R_{\mathrm{acc}}}
\newcommand{\Rbb}{R_{\mathrm{bb}}}
\newcommand{\Rsb}{R_{\mathrm{sb}}}
\newcommand{\Rmp}{R_{\mathrm{cp}}}
\newcommand{\SNR}{\mathrm{SNR}}
\newcommand{\QBER}{\mathrm{QBER}}
\begin{document}
\title{Improving the signal-to-noise ratio through postselection of coincident detections \\in anticounterfeiting and quantum key distribution}

\author{Paula Wieteska-Bajak\,\orcidlink{0009-0009-4180-4877}}
\affiliation{Aix Marseille Univ, CNRS, Centrale M\'editerran\'ee, Institut Fresnel, Marseille, France}

\author{Mohamed Hatifi\,\orcidlink{0009-0005-3368-2751}}
\affiliation{Aix Marseille Univ, CNRS, Centrale M\'editerran\'ee, Institut Fresnel, Marseille, France}

\author{Renaud Mathevet}
\affiliation{LCAR, Universit\'e de Toulouse, CNRS (UMR 5589), France}

\author{Vincent Lacombe}
\affiliation{Aix Marseille Univ, CNRS, Centrale M\'editerran\'ee, Institut Fresnel, Marseille, France}

\author{J\'er\^ome Wenger}
\affiliation{Aix Marseille Univ, CNRS, Centrale M\'editerran\'ee, Institut Fresnel, Marseille, France}

\author{Thomas Durt}
\email{thomas.durt@centrale-med.fr}
\affiliation{Aix Marseille Univ, CNRS, Centrale M\'editerran\'ee, Institut Fresnel, Marseille, France}


\begin{abstract}

In quantum communications and other photon-counting applications, reliably detecting weak optical signals against a strong noise background remains a fundamental challenge. In this paper, we develop a simple idea: when a signal is carried by pairs (or more generally multiplets) of photons emitted simultaneously, postselection of coincident detections recorded on two detectors makes it possible to improve-sometimes drastically-the Signal to Noise Ratio. Essentially, this is made possible by the fact that genuine and accidental coincidences scale differently with the width of the postselection window. As long as the window remains wide enough to retain detections from the same pair, the genuine coincidence rate is approximately unchanged. In contrast, coincidences formed from independent clicks decrease linearly as the window is narrowed. The resulting signal-to-noise ratio, therefore, increases inversely with the window width. Our approach makes it possible to reveal photon-pair signals that are otherwise hidden beneath a much stronger uncorrelated background.
We analyze here applications of this idea to quantum key distribution, metrology,  and anti-counterfeiting. 

\end{abstract}

\maketitle

\section{Introduction.}

\begin{figure}[t]
    \centering
    \includegraphics[width=0.7\columnwidth]{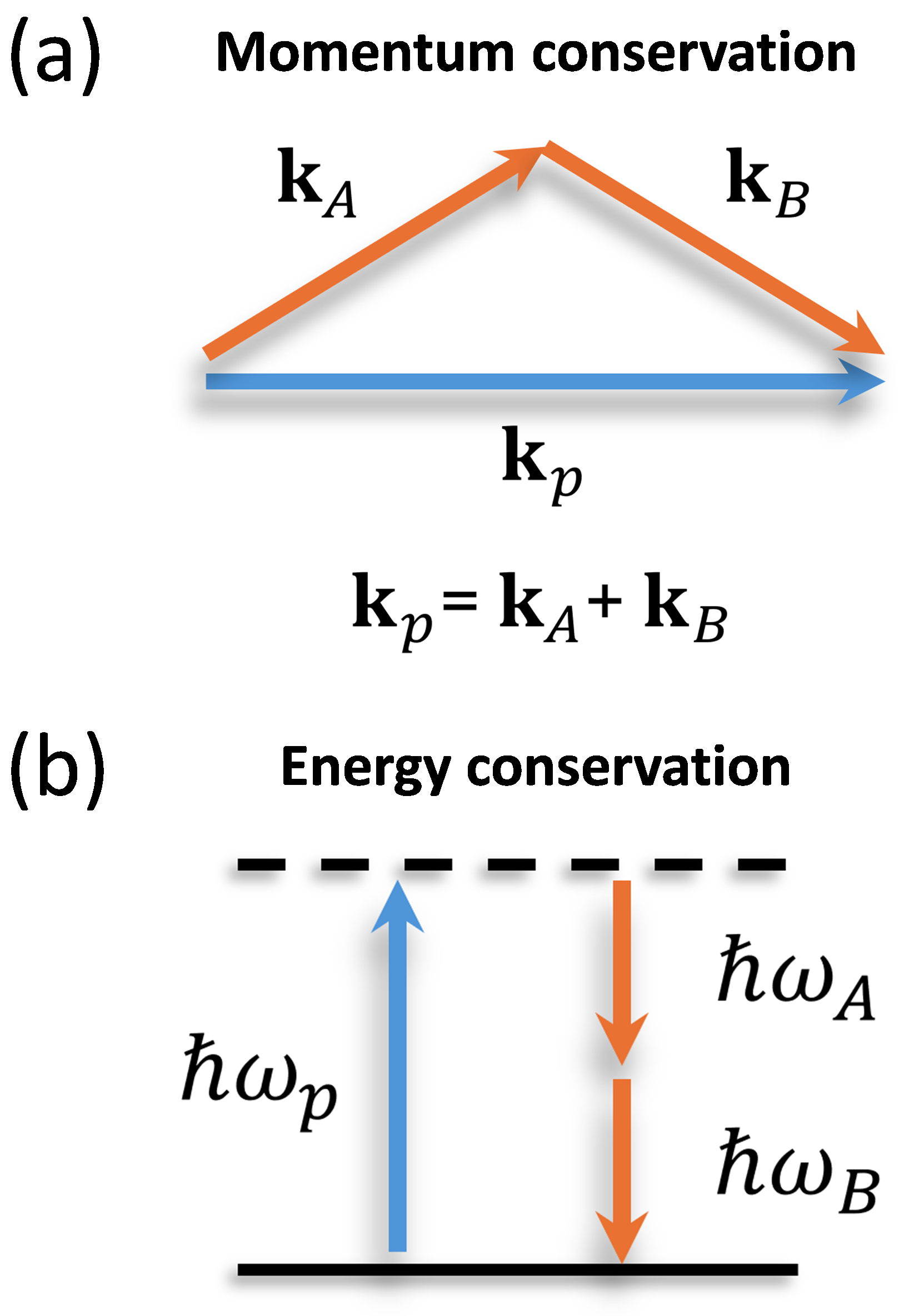}
\caption{Schematic representation of spontaneous parametric down-conversion (SPDC). A pump photon is converted into two lower-energy photons, denoted $A$ and $B$. (a) Momentum conservation (phase-matching condition), $\mathbf{k}_p=\mathbf{k}_A+\mathbf{k}_B$. (b) Energy conservation, $\omega_p=\omega_A+\omega_B$. Here, $\mathbf{k}_j$ and $\omega_j$ denote the wave vector and angular frequency of photon $j$, with $j=p,A,B$ and $p$ identifying the pump.}
    \label{fig:scheme_1}
\end{figure}

As a consequence of the projection postulate, it is most often impossible to measure a quantum system without disturbing its state. In particular, if information is encoded in complementary degrees of freedom, only an authorized user who knows the encryption basis has full access to this information, consistent with Heisenberg's uncertainty principle. \rmedit{The no-cloning theorem also prevents perfect copies of an unknown quantum state} \cite{WoottersZurek1982}. These concepts paved the way for anti-counterfeiting and quantum key distribution (QKD), which were respectively conceived by Wiesner in the 70s (though it remained unpublished until 1983 \cite{Wiesner1983}) and by Bennett and Brassard in 1984 \cite{BennettBrassard1984}. Wiesner proposed the idea of quantum money to create unforgeable banknotes. Bennett and Brassard proposed encoding a random binary key in complementary bases to ultimately serve as a one-time pad. In both cases, the support of the information is a single qubit, and the bases of encryption are so-called mutually unbiased bases (which are associated with orthogonal directions on the Bloch sphere representation of the qubit Hilbert space). In 1991, Artur Ekert proposed a two-qubit scheme in which encryption bases are chosen by two authorized users, Alice and Bob, who share a maximally entangled two-qubit state \cite{Ekert1991}. In optical implementations, these qubits are commonly encoded in photon pairs generated (see Fig \ref{fig:scheme_1}.) via a nonlinear process such as spontaneous parametric down-conversion \cite{Kwiat1999}. The use of photon pairs introduces an additional experimental task, since Alice and Bob must identify which detection events originate from the same emission process. This identification is usually performed through temporal postselection \cite{BurnhamWeinberg1970,Holloway2013}. After compensating for the relative propagation delay between the two detection channels, two clicks are assigned to the same pair when their time tags fall within a coincidence window of width $\tps$, which we refer to as the postselection window.
Genuine pair detections are retained through this temporal criterion, but independent clicks can also satisfy it by chance, producing accidental coincidences between detections that do not originate from the same emitted pair. Such events can arise from detector dark counts, ambient light, fluorescence in the optical setup, or photons from different emitted pairs \cite{Hadfield2009,Eisaman2011}.
\\ \indent The central observation of this work follows from the different dependence of genuine and accidental coincidences on the postselection window (see ~\ref{sec:ideal}). The probability of detecting a genuine pair within a window is proportional to $\tps$. In contrast, the probability of recording simultaneously one independent background click in each channel is proportional to $\tps^2$. Since the number of windows per unit time scales as $1/\tps$, the genuine coincidence rate remains independent of the window width, while the accidental coincidence rate increases linearly with it. Temporal postselection can therefore recover a correlated signal even when its contribution to every single record is much weaker than the background. We examine this principle in two settings: the first is entanglement-based quantum key distribution, in which accidental coincidences produce random outcomes, thereby increasing the quantum bit error rate. The second is optical marking (with applications in anticounterfeiting), where a small number of pair-emitting centers form a correlation marker whose presence \rmedit{may remain hidden in the background noise and be measurable through postselected coincidences. This is the "needle-in-a-haystack" approach: temporal correlations reveal a weak quantum-light marker in a strong background.}
\\ \indent
The remainder of the paper is organized as follows. In Sec.~\ref{sec:qkd}, we introduce the postselection protocol and derive the counting formulas for single-photon and photon-pair transmission. We then examine how detector inefficiency, transmission loss, multipair emission, optical pollution, and dark counts modify the ideal behavior. Section~\ref{sec:marking} applies the same framework to correlation-based optical marking and anticounterfeiting, while Sec.~\ref{sec:experiment} presents the experimental results. 
\\
In Sec.~\ref{sec:multiplets}, we extend the scaling argument to higher-order photon correlations. Additional experimental plots are provided in the Appendix.

\section{Postselection of coincident photon pairs in quantum key distribution}
\label{sec:qkd}
We consider a source that emits photon pairs at a rate $\Rp=1/\tpair$, where $\tpair$ is the mean interval between successive emission events. As illustrated in Fig.~\ref{fig:scheme_1},
during spontaneous parametric down-conversion (SPDC), a photon of energy $h\nu$ (from now on denoted the blue photon) is sent into a non-linear medium and generates a pair of low-energy photons (from now on denoted the red photons or twin photons) of energies of the order of $h\nu/2$ \cite{Kwiat1999}.

One photon propagates toward Alice, and the other toward Bob, and the total probability that they produce registered clicks is $\eta_A=\eta_{\mathrm{trans}}^A\eta_{\mathrm{det}}^A$ and $\eta_B=\eta_{\mathrm{trans}}^B\eta_{\mathrm{det}}^B$, respectively. These quantities include both propagation losses and detector inefficiencies. The registered background rates at the two receivers are denoted by $R^{bn}_A$ and $R^{bn}_B$; they represent the total rates over the detectors used at each side and may include dark counts, ambient photons, \rmedit{and other random contributions.} For symmetric configurations, we write $R^{bn}_A=R^{bn}_B=1/\tbn$. The detection time tags are divided into windows of width denoted $\tps$. \rmedit{An event is retained only when exactly one click is registered at Alice's side and exactly one at Bob's side.}

In contrast, windows containing multiple clicks at either receiver are discarded. Any known difference in propagation times between the two channels is removed by an appropriate delay imposed at the time tagger before this selection is applied, so two detections are considered simultaneous when they fall within the same compensated window. 

In what follows, we will consider that the number of pair-emission events within a fixed postselection window is modeled by a Poisson distribution \cite{Neumann2021}. Other distributions could be considered (e.g., thermal or geometric distributions), but alternative choices of distribution do not substantially alter our analysis; therefore, in the present paper we will, once and for all, assume that the distribution of pairs is Poissonian. The mean number of emitted pairs per window is thus

\begin{equation}
\mu=\Rp\tps=\frac{\tps}{\tpair}.
\label{eq:mu}
\end{equation}
Accordingly, the probabilities of producing one and two pairs are $P_1=\mu e^{-\mu}$ and $P_2=\mu^2e^{-\mu}/2$. In the sparse regime, where $\mu\ll1$, these expressions reduce to $P_1\simeq\tps/\tpair$ and $P_2\simeq(\tps/\tpair)^2/2$. The same requirement must hold for the background processes, and the analysis below, therefore, assumes
\begin{equation}
\Rp\tps\ll1,
\qquad
R^{bn}_A\tps\ll1,
\qquad
R^{bn}_B\tps\ll1.
\label{eq:sparseconditions}
\end{equation}

These inequalities set an upper range on useful postselection windows. In practice, the lower useful range of the postselection window is determined by the detector timing jitter, time-tagger resolution, synchronization uncertainty, and the dependence of the detector response on the background count rate \cite{Kuniyil2026}. For the QKD analysis, we consider only the events retained after the measurement bases have been compared (sifting). We denote by $R_{\mathrm{correct}}$ the rate of retained events that exhibit the maximal correlation expected for the chosen pair of bases, and by $R_{\mathrm{error}}$ the rate of retained events that give the opposite outcome. Accidental coincidences are uncorrelated and therefore contribute equally, on average, to these two rates. We characterize the quality of the postselected data through the correct-to-error ratio
\begin{equation}
\SNR_{\mathrm{QKD}}=\frac{R_{\mathrm{correct}}}{R_{\mathrm{error}}},
\label{eq:SNRdefinition}
\end{equation}
which measures the quality of the retained bits. The corresponding quantum bit error rate is 
\begin{equation}
\QBER=\frac{R_{\mathrm{error}}}{R_{\mathrm{correct}}+R_{\mathrm{error}}}
=\frac{1}{1+\SNR_{\mathrm{QKD}}}.
\label{eq:QBERdefinition}
\end{equation}

\subsection{Comparison between single-photon and pair protocols}
\label{sec:ideal}
Let us first consider the effect of temporal postselection in an ideal situation: perfect transmission and detection, with no intrinsic state-preparation error (maximal entanglement ~\ref{sec:intrinsic} Eq.~\eqref{eq:Werner}). 
\subsubsection{Single-photon protocol}

Let $\tau_{\mathrm{single}}$ denote the mean interval between successive signal photons sent to Bob, so that the signal rate is $1/\tau_{\mathrm{single}}$. During a window of width $\tps$, the probabilities of receiving a signal photon and a background click are, respectively, $\tps/\tau_{\mathrm{single}}$ and $\tps/\tbn$. In the ideal model, a signal photon produces the correct bit, whereas a background click is randomly assigned to either of the two possible outcomes. The correct and erroneous retained-bit rates are therefore
\begin{equation}
R_{\mathrm{correct}}
=\frac{1}{\tau_{\mathrm{single}}}+\frac{1}{2\tbn},
\qquad
R_{\mathrm{error}}
=\frac{1}{2\tbn}.
\label{eq:singlecorrecterror}
\end{equation}
The factor of $2$ appears because Bob will interpret a background click as a bit 0 or 1 with equal probability. Using Eq.~\eqref{eq:SNRdefinition}, we obtain
\begin{equation}
\SNR_{\mathrm{single}}
=1+2\frac{\tbn}{\tau_{\mathrm{single}}}
\simeq2\frac{\tbn}{\tau_{\mathrm{single}}},
\label{eq:OnePhoton}
\end{equation}
where the final form applies when the signal rate is much larger than the background rate, which is the case in the BB84 protocol, for which security imposes that the quantum bit error rate is upper-bounded by approximately 11 $\%$ if we consider coherent attacks and 14.64 $\%$ in the case of cloning-based individual attacks \cite{Cerf2002}.

Both the signal and background probabilities are proportional to $\tps$, so their ratio is independent of the window width. Therefore, modifying the width of the postselection window does not improve the retained-bit fidelity in this case. It is worth noting at this level that we consider a transmission protocol for which we cannot predict in advance the temporal window within which photons sent from Alice to Bob will arrive and/or be detected. This is achieved in practice by using a strongly attenuated laser beam as a source. QKD protocols involving deterministic single-photon guns, and in particular the case where single photons are heralded, will be considered in another section~\ref{sec:heralded}.
\subsubsection{Pair protocol}

Here, we show that the situation changes when maximally entangled photon pairs carry the signal. In the ideal limit considered here, errors due to mixed signal--background coincidences do not survive the postselection process: both photons of every emitted pair are detected, so an additional background click produces multiple clicks at the level of one receiver, and the corresponding window is discarded. More generally, we will neglect in the rest of the paper situations in which three or more photons are present during the same temporal window, because their contribution is small; it can be considered as a second-order effect which disappears in the limit where the signal is sparse Eq.~\eqref{eq:sparseconditions} . Mixed signal--background coincidences must be taken into account in non-ideal situations (e.g. due to lossy transmission and/or to dark count in inefficient detectors); however, as discussed in the next subsection. 

To leading order in the sparse regime, the probability of producing one pair inside a window is $\tps/\tpair$, so the genuine coincidence rate (from now on denoted $R_{twinp}$) remains
\begin{equation}
R_{twinp}=\frac{1}{\tpair}=\Rp.
\label{eq:Rtrueideal}
\end{equation}
By contrast, an accidental coincidence requires one independent background click at each receiver. Its probability is $(\tps/\tbn)^2$, and the corresponding rate is
\begin{equation}
\Rbb
=\left(\frac{\tps}{\tbn}\right)^2\frac{1}{\tps}
=\frac{\tps}{\tbn^2}.
\label{eq:Rbbideal}
\end{equation}
A genuine pair yields the expected correlated outcome, whereas an accidental coincidence contributes equally to both correct and erroneous bits. Hence
\begin{equation}
R_{\mathrm{correct}}
=\frac{1}{\tpair}+\frac{\Rbb}{2},
\qquad
R_{\mathrm{error}}
=\frac{\Rbb}{2},
\label{eq:paircorrecterror}
\end{equation}
which yields
\begin{equation}
\SNR_{\mathrm{pair}}
=1+\frac{2\tbn^2}{\tps\tpair}
.
\label{eq:TwoPhoton}
\end{equation}
The physical origin of the enhancement is thus explicit: genuine coincidences require one pair-emission event and therefore have an approximately window-independent rate, whereas accidental coincidences require two independent detections and grow linearly with $\tps$. Narrowing the postselection window consequently improves the correct-to-error ratio to $1/\tps$, provided the sparse-counting conditions remain satisfied Eq.~\eqref{eq:sparseconditions}.
\subsubsection{Illustration: some orders of magnitude.}

To illustrate the contrast between the two scalings, we take equal background rates $R^{bn}_A=R^{bn}_B=\SI{100}{s^{-1}}$. For a single-photon signal received at $\SI{1e4}{s^{-1}}$, Eq.~\eqref{eq:OnePhoton} gives $\SNR_{\mathrm{single}}\simeq2\times10^2$. On the other hand, for a pair source with $\Rp=\SI{100}{s^{-1}}$ and a postselection window $\tps=\SI{1e-7}{s}$, the accidental coincidence rate is
\begin{equation}
\Rbb=R^{bn}_A R^{bn}_B\tps=\SI{1e-3}{s^{-1}},
\label{eq:Rbbexample}
\end{equation}
and Eq.~\eqref{eq:TwoPhoton} yields $\SNR_{\mathrm{pair}}\simeq2\times10^5$. Reducing the window to $\SI{1e-8}{s}$ increases this ratio to approximately $2\times10^6$. Even when the pair-production rate is lowered to $\SI{1}{s^{-1}}$, a window of $\SI{1e-7}{s}$ still gives $\SNR_{\mathrm{pair}}\simeq2\times10^3$. The same mechanism becomes particularly clear when the pair signal is weaker than the background in each channel. For $\tps=\SI{1e-8}{s}$, the accidental coincidence rate is $\SI{1e-4}{s^{-1}}$. A source producing $\SI{1e-2}{pair\,s^{-1}}$ then generates genuine coincidences at a rate one hundred times larger than the accidental rate. However, its contribution to each single record is $10^4$ times smaller than the background. The pair source is therefore almost invisible in the individual detection channels, yet remains clearly identifiable through its temporal correlations. This is our main source of inspiration for the "needle-in-a-haystack" approach to optical marking developed in Sec.~\ref{sec:marking}.
\\
\subsection{\rmedit{Transmission losses and detector efficiencies}}
\label{sec:loss}

\rmedit{In realistic settings, transmission losses and finite detector efficiencies reduce the probability} that both photons of an emitted pair are registered. With the total detection probabilities introduced in \rmedit{Sec.~\ref{sec:qkd}}, the true or genuine coincidence rate becomes
\begin{equation}
R_{twinp}=\Rp\eta_A\eta_B.
\label{eq:Rtruegeneral}
\end{equation}
For equal detector efficiencies and lossless channels, this gives $R_{twinp}=\eta_{\mathrm{det}}^2/\tpair$, while perfect detectors in lossy channels give $R_{twinp}=\eta_{\mathrm{trans}}^A\eta_{\mathrm{trans}}^B/\tpair$. Independent registered backgrounds at Alice's and Bob's receivers produce accidental coincidences at the rate
\begin{equation}
\Rbb=R^{bn}_A R^{bn}_B\tps.
\label{eq:Rbbgeneral}
\end{equation}
where \rmedit{$\eta_A=\eta_{\mathrm{det}}^A\eta_{\mathrm{trans}}^A$ and $\eta_B=\eta_{\mathrm{det}}^B\eta_{\mathrm{trans}}^B$.} Writing this contribution in terms of the measured rates $R^{bn}_A$ and $R^{bn}_B$ is important because environmental background noise and detector dark counts do not respond to detector efficiency in the same way. For example, if the background consists only of noisy single photons with rate $1/\tbn$ at each detector, then $R^{bn}_A=R^{bn}_B=\eta_{\mathrm{det}}/\tbn$ and $\Rbb=\eta_{\mathrm{det}}^2\tps/\tbn^2$. Dark counts, by contrast, must be included directly in $R^{bn}_A$ and $R^{bn}_B$. 

\rmedit{These inefficiencies also permit accidental coincidences absent from the ideal model:} a photon from a generated pair may be registered at one receiver while its partner is lost, with an independent background click supplying the second accepted event. \rmedit{Including the symmetric process, we obtain}
\begin{equation}
\Rsb=\Rp\tps\left[
\eta_A(1-\eta_B)R^{bn}_B
+\eta_B(1-\eta_A)R^{bn}_A
\right],
\label{eq:Rsb}
\end{equation}

A second possibility arises when two different pairs are emitted within the same window. One detected photon may then originate from each pair, while the two partners are lost or undetected. Such photons are cousins, not twins, and thus not entangled with each other. When two pairs are emitted by the source inside the same temporal window, and only one photon arrives at each side, \rmedit{there are two ways to form accidental coincidences. Therefore, to leading order} in $\Rp\tps$, this cross-pair contribution is thus
\begin{equation}
\Rmp=\Rp^2\tps\eta_A\eta_B
(1-\eta_A)(1-\eta_B).
\label{eq:Rmp}
\end{equation}
Within the sparse-counting approximation, the total accidental rate is therefore
\begin{equation}
\Racc=\Rbb+\Rsb+\Rmp,
\label{eq:Racc}
\end{equation}

where $\Racc$, $\Rbb$, $\Rsb$ and $\Rmp$ respectively denote the total rate of accidental pair detections,  of simultaneous detection of a background single photon at each side, of a signal photon at one side and a background photon at the other side and, finally, of two (red) signal photons that were not produced by the same blue (pump) photon.  All three terms remain proportional to $\tps$, but their relative weights respond differently on the source brightness, background level, and total detection efficiencies. 
All other postselected pairs correspond to the effective signal, characterized by the rate $R_{twinp}$.
\subsection{Intrinsic state error and retained-bit quality}
\label{sec:intrinsic}

Temporal postselection suppresses coincidences formed from unrelated detections, but it cannot correct errors already present in the photon-pair state. It is therefore useful to separate two physically distinct contributions to the retained-bit error rate. \rmedit{The first arises from imperfect correlations within genuinely detected pairs. The second arises} from accidental coincidences between clicks that do not belong to the same pair. We model imperfect pair preparation by the Werner state \cite{Werner1989}
\begin{equation}
\rho=\rmedit{(1-\epsilon)\ket{\Psi}\!\bra{\Psi}+\epsilon\frac{\mathbb I_4}{4}},
\label{eq:Werner}
\end{equation}
where $\ket{\Psi}$ is the maximally entangled target state and $\epsilon$ \rmedit{quantifies the admixture of the maximally mixed state. The limit $\epsilon=0$ describes a perfectly prepared entangled state, whereas $\epsilon=1$ gives a completely mixed two-qubit state.}

We consider a measurement basis in which $\ket{\Psi}$ produces the expected correlation with unit probability. The ideal component, therefore, contributes only to the correct retained-bit rate, with a contribution \rmedit{$(1-\epsilon)R_{twinp}$}. By contrast, the maximally mixed component assigns equal probability to the four possible joint outcomes. Two of these outcomes obey the expected correlation, and two violate it, so this component contributes equally to correct and erroneous bits. The same reasoning applies to the accidental coincidences collected in $\Racc$, since their outcomes are uncorrelated in the selected basis. The retained bit rates are therefore
\begin{align}
R_{\mathrm{correct}}
&=\rmedit{(1-\epsilon)\Rtrue+\frac{\epsilon}{2}\Rtrue}+\frac{1}{2}\Racc,
\label{eq:Rcorrectgeneral}\\
R_{\mathrm{error}}
&=\rmedit{\frac{\epsilon}{2}\Rtrue}+\frac{1}{2}\Racc.
\label{eq:Rerrorgeneral}
\end{align}
The corresponding correct-to-error ratio is
\begin{equation}
\SNR_{\mathrm{QKD}}
=
\frac{
\rmedit{(1-\epsilon)\Rtrue+\frac{1}{2}\left[\epsilon\Rtrue+\Racc\right]}
}{
\rmedit{\frac{1}{2}\left[\epsilon\Rtrue+\Racc\right]}
},
\label{eq:SNRgeneral}
\end{equation}
while the total retained-bit rate is
\begin{equation}
R_{\mathrm{correct}}+R_{\mathrm{error}}=\Rtrue+\Racc.
\label{eq:Rretainedtotal}
\end{equation}
The quantum bit error rate consequently becomes
\begin{equation}
\QBER=
\frac{\rmedit{\epsilon\Rtrue+\Racc}}
{2(\Rtrue+\Racc)}.
\label{eq:QBERgeneral}
\end{equation}
The term \rmedit{$\epsilon\Rtrue$} represents errors carried by genuine detected pairs because their prepared state is imperfect. In contrast, the term $\Racc$ represents errors introduced by background--background, signal--background, and cross-pair coincidences. Both contributions are weighted by 1/2 because a randomly retained event is erroneous with probability 1/2. Several limiting cases clarify the physics and give a pedagogical understanding. For a perfect source and negligible accidentals, \rmedit{$\epsilon=0$} and $\Racc\to0$, so that $\QBER\to0$. If the state is perfect but accidental coincidences dominate, then
\begin{equation}
\QBER=
\frac{\Racc}{2(\Rtrue+\Racc)}
\longrightarrow\frac{1}{2}
\qquad
(\Racc\gg\Rtrue),
\end{equation}
which is the completely random-bit limit.

Most importantly, narrowing the postselection window suppresses $\Racc$ through Eq.~\eqref{eq:Racc}, but it does not modify the state parameter $\epsilon$. Hence, once accidental coincidences have been made negligible, the QBER saturates at the intrinsic error floor
\begin{equation}
\QBER_{\mathrm{int}}
=
\lim_{\Racc\to0}\QBER
=
\rmedit{\frac{\epsilon}{2}}.
\label{eq:QBERintrinsic}
\end{equation}
Temporal postselection can therefore remove errors caused by incorrect temporal associations, but it cannot restore correlations lost during state preparation, transmission, or measurement. In practice, $\epsilon$ should therefore be inferred from the measured correlation visibility (as shown explicitly in Table \ref{tab:theory_experiment}) or derived from a microscopic noise model. Its dependence on propagation distance is not universal, since different channels may introduce depolarization, dephasing, mode mismatch, or basis-dependent errors.
\subsection{Illustration: Moon-Earth QKD.}
\label{sec:moonmain}
In fibered QKD, the transmission decreases exponentially with the distance between the source and the detectors \cite{Pirandola2017}. As the dark-count rate remains approximately constant while the detected signal decreases, the SNR decreases with increasing transmission distance. This loss is a major limitation for conventional single-photon and entanglement-based fiber QKD, for which record distances are typically of the order of a few hundred kilometers \cite{Boaron2018,Neumann2022}. Specialized architectures can extend this range substantially; in particular, twin-field QKD has been demonstrated over more than 1000 km of fiber \cite{Liu2023}. In the case of free-space transmission, however, it is possible to realize QKD over much longer distances \cite{DurtSchneider2026}. In particular, satellite-to-earth communication over 1200 kilometers was achieved by a Chinese team several years ago \cite{Yin2017}, which, to date, constitutes the longest distance over which \rmedit{violation of Bell's inequalities} has been demonstrated. \rmedit{One reason for the success of these free-space techniques is that atmospheric density, and hence the amount of matter causing absorption and scattering, decreases with altitude.} A theoretical study has also investigated Bell tests and entanglement distribution through losses representative of an Earth--Moon link \cite{Cao2018}. Accessory, violating Bell's inequalities over the Moon-Earth distance (390,000 kilometers) would allow a gain of a factor of 300 over the distance of entanglement. Such an experiment would constitute a new fundamental test of Quantum Physics and allow for tighter constraints on alternative theories and interpretations \cite{DurtSchneider2026}.

\rmedit{For such transmission distances}, which constitute a regime of extreme loss, the competition between background and multipair events becomes striking\footnote{As noted in Ref.~\cite{Holloway2013}, the cross-pair contribution gives rise to the familiar trade-off between increasing the source brightness and preserving the quality of the postselected correlations in entanglement-based QKD.} as we show now. Inspired by Ref. \cite{Cao2018}, let us consider a source on a satellite that sends entangled pairs to Alice and Bob, with Alice on the Moon and Bob on Earth. Following  Ref. \cite{Cao2018}, we consider an ultra-bright source of pairs with $\Rp=10^{11}\,\mathrm{s}^{-1}$. To simplify the treatment, we will also assume here that the registration probabilities are equal and of the order of  $\eta_A=\eta_B=10^{-5}$ (this differs from the proposal of Ref. \cite{Cao2018}, but introducing asymmetric values for $\eta_A$ and $\eta_B$ does not drastically modify our analysis). The true coincidence rate is then
\begin{equation}
\Rtrue=\Rp\eta_A\eta_B=\SI{10}{s^{-1}}.
\label{eq:moontrue}
\end{equation}
We estimate the background rates $R^{bn}_A$ and $R^{bn}_B$ to be of the order of $\SI{100}{s^{-1}}$ , a representative low-noise value for high-quality SPAD systems \cite{Wayne2021,Hadfield2009}. As shown in Ref.\cite{Cao2018}, because $\Rtrue=\SI{10}{s^{-1}},$ it is sufficient for Alice and Bob to randomly choose their basis of detection at the same rate to realize QKD. Now, this is clearly not sufficient. Supplementary to Alice's and Bob's random choice, it is \rmedit{necessary for Alice and Bob to record the time tags of their clicks first and then perform postselection} in accordance with the rules described at the beginning of this paper. To be able to differentiate twin photons from cousin photons, a postselection window of at most, say, $\tps=\SI{1e-12}{s}$ is thus required.  Eqs.~\eqref{eq:Rbbgeneral}--\eqref{eq:Rmp} give
\begin{equation}
\Rbb=\SI{1e-8}{s^{-1}},
\quad
\Rsb\simeq\SI{2e-4}{s^{-1}},
\quad
\Rmp\simeq\SI{1}{s^{-1}}.
\label{eq:moonaccidentals}
\end{equation}
The narrow window makes both background contributions negligible, but it does not eliminate coincidences formed from photons belonging to different emitted pairs. Because this contribution grows quadratically with the source rate, it dominates the accidental counts in the present high-brightness regime. For an ideal prepared state, \rmedit{$\epsilon=0$}, Eq.~\eqref{eq:QBERgeneral} yields
\begin{equation}
\QBER\simeq4.5\%.
\label{eq:moonqber}
\end{equation}
Here, the mean number of generated pairs per window is $\mu=\Rp\tps=0.1$. Note that a  $\SI{1}{ps}$ postselection window is useful only if the source, detectors, synchronization, propagation, compensation, and time tagging preserve true coincidences on that timescale. Our analysis has the merit of emphasizing the technological performance required to realize QKD over such extreme transmission distances.

\subsection{Comparison with heralded single-photon strategies}
\label{sec:heralded}
We now compare pair-based protocols with single-photon protocols in which one of the photons emitted by the source (say, Alice's photon) is detected immediately after emission. At the same time, the second one is sent to Bob.  A pair source used in a heralded configuration provides a detection event that announces the photon's emission time as it travels through the communication channel, allowing background clicks outside the heralding interval to be rejected \cite{Eisaman2011}. Alternatively, Alice could choose to send an attenuated laser pulse with, on average, one photon per pulse, at a pulse rate equal to $1/\tps$, using decoy-state methods to detect photon-number-splitting attacks \cite{Durt1999Comment} associated with the multiphoton components of weak coherent pulses \cite{LoMaChen2005, Qiu2025}. The main advantage of such single-photon protocols where the source is a (nearly) deterministic single-photon gun is that the effective bit rate is quite higher than in the case of symmetric QKD {\it \`a la} Ekert, where the transmission factor is squared. For instance, in the case of Moon-Earth transmission, Alice could send bits to Bob at a rate $\Rp \eta_B=\SI{1e6}{s^{-1}}$, to compare with the effective rate reached in the case of symmetric QKD {\it \`a la} Ekert ($\SI{10}{s^{-1}}$).
\\
Let us now assume that Bob and Charles, both on Earth, wish to refresh a shared one-time pad key using photons emitted by a source operated by Alice on a satellite or on the Moon. If Alice acts as a trusted relay and has access to the distributed key, its secrecy depends on her honesty. A separate concern arises if Alice also supplies measurement devices whose operation Bob and Charles cannot verify. Such devices could report correlated outcomes generated by a software program containing preregistered shared "hidden variables" known to Alice \cite{Brans1988,Durt1997,Durt2001Comment,Aktas2015}. Bob and Charles could then obtain an apparent key whose contents are already known to her. Observing these correlations alone would therefore be insufficient to establish security. Entanglement-based QKD can accommodate an untrusted source when the users' measurement devices are trusted. If those devices are also untrusted, device-independent security requires assessing a Bell violation without the relevant loopholes, along with the assumptions and security analysis of the chosen protocol.
\\
In the case of QKD {\it \`a la} Ekert, the violation of Bell's inequalities can support security claims, subject to the protocol's assumptions and its security analysis. Moreover, in the ideal case of maximally entangled pairs, the owner of the source has no information about the key shared by Bob and Charles, by virtue of the monogamy of entanglement. This property is closely related to the idea of device independence \cite{Acin2007,Lu2026}. It is, however, beyond the scope of the present paper to develop this analysis.


\section{Pair-emitting markers for optical authentication}
\label{sec:marking}

\subsection{Correlation marking in the needle-in-a-haystack regime}

Detecting weak optical markers in background light is one application of temporal postselection that we investigate in the ARTEMIS project \cite{ARTeMIS,Hatifi2022}. Earlier work has examined strong-coupling effects for optical marking \cite{Hatifi2022} and the storage and revival of Bell nonlocality through emitter geometry and environmental memory \cite{hatifi2026d}. Existing anti-counterfeiting labels encode information in several measurable properties. Random fluorescence patterns can identify individual labels, as demonstrated with inkjet-printed quantum dots forming flower-like structures authenticated by image recognition \cite{Liu2019}. Measurements under different illumination conditions provide further checks, including the nonlinear dependence of photoluminescence on excitation power for quantum dots encapsulated in a transparent polymer \cite{Fong2022}, and polarization-dependent patterns from fluorescent nanodiamonds authenticated using deep metric learning \cite{Wang2024}. Other markers carry information in their composition. Examples include isotope-labeled glucose added to pharmaceutical formulations and identified by mass spectrometry \cite{Felton2011}, and engineered DNA mixtures read through hybridization as fluorescent patterns on paper-based tickets \cite{Li2025DNA}.

Here, the measured signature is the temporal correlation between photons originating from the same emission event. We consider an optically excited photon-pair-emitting marker whose emission is collected into two detection channels. Selecting coincidences within a short time window reduces accidental associations between independent background detections. Provided that the window still accommodates the timing uncertainty of genuine pairs, narrowing it can reveal the marker's contribution even when the excess intensity is difficult to resolve in either singles record.

The interpretation of mixed events, however, differs from the QKD case. A coincidence formed by one marker photon and one background click carries no reliable qubit correlation and therefore contributes to the QKD error rate. In an on--off marker measurement, however, the same event remains associated with the presence of the illuminated marker and contributes to the measured coincidence excess. A false marker indication instead arises from coincidences that persist in the marker-absent reference, such as background--background events. This interpretation assumes that the background statistics remain unchanged between the marker-present and marker-absent measurements.

In anticounterfeiting, it is assumed that we excite a sample containing markers and that the emitted photons are detected in the vicinity of the sample. As always, we include the short propagation path between the sample and the detectors in the total registration probabilities $\eta_A$ and $\eta_B$. 
Moreover, as the signal is carried by the presence of a photon and not by its qubit degree of freedom (polarisation, time bins and so on), it suffices, to reveal the presence of pair emitters, to estimate the difference between the (simultaneous pair) count rate in the presence of emitters and the count rate in their absence.

The detected genuine-pair contribution is, as always
\begin{equation}
R_{\mathrm m}^{(0)}=\Rp\eta_A\eta_B,
\label{eq:Rmarkertrue}
\end{equation}
while coincidences containing one detected marker photon and one background click occur at the rate
\begin{equation}
R_{\mathrm m}^{(1)}
=\Rp\tps\left[
\eta_A(1-\eta_B)R^{bn}_B
+\eta_B(1-\eta_A)R^{bn}_A
\right].
\label{eq:Rmarkermixed}
\end{equation}
In the absence of the marker, the reference coincidence rate is
\begin{equation}
R_0=R^{bn}_A R^{bn}_B\tps.
\label{eq:Rmarkerbackground}
\end{equation}
\rmedit{We consider markers that are sufficiently dilute for their emission to remain weak. In this regime, higher-order pair emission can be neglected, and the marker-dependent excess is}
\begin{equation}
R_{\mathrm m}=R_{\mathrm m}^{(0)}+R_{\mathrm m}^{(1)},
\label{eq:Rmarkerexcess}
\end{equation}
and we quantify its visibility through the rate contrast
\begin{equation}
\mathcal C_{\mathrm m}=\frac{R_{\mathrm m}}{R_0}.
\label{eq:SNRmarker}
\end{equation}

We adopt here the definition of $\tbn$ introduced in Sec.~\ref{sec:qkd} through the registered background rate. Therefore, the detected background rates are  $R^{bn}_A=1/\tau^A_{\mathrm{bn}}$ and $R^{bn}_B=1/\tau^B_{\mathrm{bn}}$.
Fluorescence clicks and environmental noise may be included through the general registered rates $R^{bn}_A $ and $R^{bn}_B$, having in mind that they do not follow the optical-efficiency scaling in $\eta_A$ and $\eta_B$ as the dark count.
Hence,

\begin{equation}
R_0=\frac{\tps}{\tau^A_{\mathrm{bn}}\tau^B_{\mathrm{bn}}}.
\label{eq:R0special}
\end{equation}
Substituting these rates into Eq.~\eqref{eq:SNRmarker} gives
\begin{equation}
\mathcal C_{\mathrm m}
=
\eta_A\eta_B\frac{\tau^A_{\mathrm{bn}}\tau^B_{\mathrm{bn}}}{\tps\tpair}
+
\frac{1}{\tpair}
\left[\tau^A_{\mathrm{bn}}\eta_A(1-\eta_B)+\tau^B_{\mathrm{bn}}\eta_B(1-\eta_A)\right].
\label{eq:SNRmarkerExplicit}
\end{equation}

The first term originates from genuine marker pairs and displays the same inverse-window enhancement as the pair contribution in the QKD analysis. The second term arises from mixed marker--background events. Because both the mixed contribution and the marker-absent reference rate are proportional to $\tps$, their ratio is independent of the window width.

The agreement between the theoretical prediction of Eq.~\eqref{eq:SNRmarkerExplicit} and the experimental results is examined in Sec.~\ref{sec:experiment1} and in the appendix. The relative deviation between the theoretical and experimental values is approximately \(10\%\) for the contribution associated with the first term of Eq.~\eqref{eq:SNRmarkerExplicit}, as discussed in connection with Eq.~\eqref{eq:Cind}, and approximately \(4.1\%\) for the mixed marker–background contribution described by the second term, as shown by Eqs.~\eqref{eq:bcrosspredvalue} and~\eqref{eq:bcrossfit}.


\section{Extension to photon multiplets}
\label{sec:multiplets}
The temporal advantage becomes stronger when an accepted event requires more than two correlated detections \cite{HigherOrderReview2022}. Let us now consider a source that produces simultaneous $N$-photon events at a detection rate $R_N$, with one photon registered in each of the $N$ channels. The probability of a genuine event within a window of width $\tps$ is $R_N\tps$. By contrast, an accidental $N$-fold coincidence requires, in first approximation (neglecting here corrections due to departures from ideality), one independent background click in every channel. For registered single photons background rates $R^{bn}_1,\ldots,R^{bn}_N$, its probability is
\begin{equation}
P_{\mathrm{acc}}^{(N)}
=\left(\prod_{j=1}^{N}R^{bn}_j\right)\tps^N,
\label{eq:PNacc}
\end{equation}
and the corresponding accidental rate is
\begin{equation}
R_{\mathrm{acc}}^{(N)}
=\left(\prod_{j=1}^{N}R^{bn}_j\right)\tps^{N-1}.
\label{eq:RNacc}
\end{equation}
As long as the postselection window retains the genuine multiplet, the ratio between genuine and accidental event rates therefore scales as
\begin{equation}
\SNR_N
=\frac{R_N}{R_{\mathrm{acc}}^{(N)}}
\propto\frac{1}{\tps^{N-1}}.
\label{eq:SNRN}
\end{equation}
A true triplet source\footnote{In practice, the observation of this higher-order enhancement is constrained by detector efficiency, dark counts, timing jitter, and finite acquisition time. These limitations have been analyzed explicitly for noisy $N$-fold coincidence measurements and photon-triplet sources \cite{ChenFainman2024}.} consequently yields a $1/\tps^2$ enhancement, while a genuine four-photon source yields a $1/\tps^3$ enhancement. A fourfold event arising from two independent photon pairs scales differently because the desired signal itself requires two emission events within the same time window. Its probability is proportional to $(\Rp\tps)^2$, whereas four independent background detections occur with probability proportional to $\tps^4$. After conversion to rates, their ratio scales as $1/\tps^2$, rather than the $1/\tps^3$ scaling of a genuine four-photon emission event. More generally, the temporal advantage is determined by the difference between the number of independent events required to generate the desired signal and the number required to reproduce it from an uncorrelated background.

Higher-order postselection can therefore provide increasingly strong rejection of an independent background. It is beyond the scope of the present paper to study in depth the final gain in SNR. One should, however, keep in mind that the potential gain for photon multiplets with $N$ larger than 2 is accompanied by a rapidly decreasing detected signal, since the registration probability is the product of the creation, transmission, and detection efficiencies of all participating photons. Candidate multiplet sources include cascaded radiative transitions and higher-order nonlinear optical processes \cite{Hubel2010,Corona2011,Pan2012}, which are generally characterized by very small coupling constants; consequently, we expect the creation rate to decrease drastically for photon triplets, quadruplets, and so on. 

\begin{figure*}[t]
\centering
\includegraphics[width=.95\textwidth]{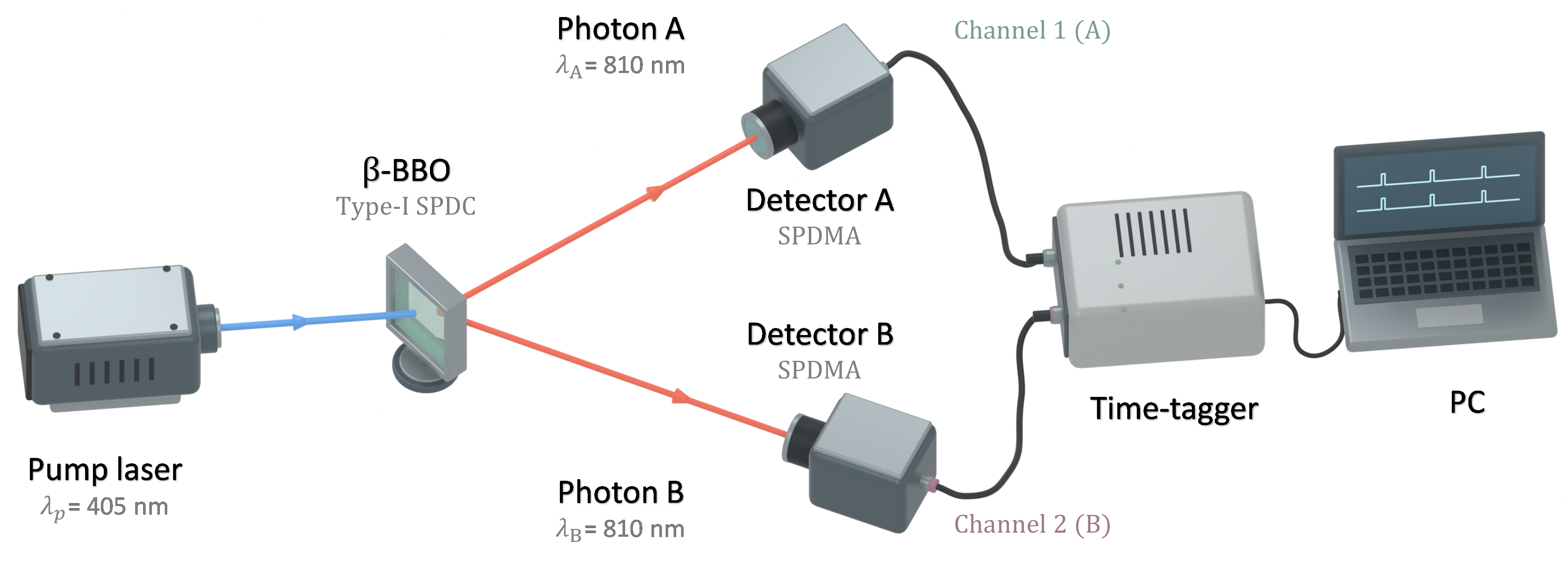}
\caption{Simplified schematic of the experimental setup used to generate and detect temporally correlated photon pairs. A continuous-wave laser at \(405\,\mathrm{nm}\) pumps a type-I (\(\beta\)-BBO) crystal, producing photon pairs near \(810\,\mathrm{nm}\). The photons are collected in two separate arms and detected by single-photon counting modules A and B. Electrical pulses from the detectors are recorded on separate channels of a time-tagger, and the acquisition data are transferred to a computer for temporal postselection analysis.}
\label{fig:experimental_setup1}
\end{figure*}

\section{Experimental demonstration of optical marking with pairs of simultaneous photons.}
\label{sec:experiment1}

\subsection{Experimental setup and measurement protocol}
To implement our ideas on marking and anti-counterfeiting, we used the Thorlabs EDU-QOP1 pedagogical kit to generate pairs of identical photons that are uncorrelated in polarization but correlated in time \cite{ThorlabsQOP}. 
The experimental arrangement is shown in Fig.~\ref{fig:experimental_setup1}.  The down-converted photons are emitted near \(810\,\mathrm{nm}\) into two spatially separated arms.

The electrical pulses produced by the two detectors are sent to separate inputs of a time tagger, denoted channels \(A\) and \(B\), and the resulting detection times are transferred to a computer for post-processing. The measurement consists of recording the singles and coincidence rates reported by the software as the width \(\tps\) of the temporal postselection window is varied from \SI{3}{ns} to \SI{300}{\micro\second}. Before applying the coincidence criterion, the relative delay between the two detection channels is compensated so that detections originating from the same photon pair are centered at zero relative delay. A detection in channel \(A\) is then associated with a detection in channel \(B\) when the two compensated time tags fall within the selected window. \rmedit{It should be, however, noted that the pedagogical setup has several limitations.} For instance, the dark count rate is high compared to more sophisticated and precise devices. \rmedit{The quantitative fits presented below use window widths of \SI{3}{ns} and above.} Moreover, we had no access to statistics on the timing of individual clicks, which remained hidden in the ``black box'' of the time-tagger. It was thus not possible to apply the strategy described in section \ref{sec:qkd}, that is to say, to retain a window only when exactly one click is registered at Alice's side, and exactly one click is registered at Bob's side. This is because the kit's time tagger systematically counts and integrates, for each detector, the number of successive clicks within a given window. When there is at least one click in each detector within the same temporal window, it also accumulates the number of clicks across both detectors. Finally, the software delivers the rates of single clicks, as well as the rates of coincidences established in this manner. \rmedit{At long windows, the software counting procedure differs from the coincidence protocol modeled in Sec.~\ref{sec:qkd}; its results cannot be interpreted by the sparse-window equations. However, for short windows, multiple clicks per channel within one window are rare. The measured rates approach the predictions of that model.} Now, it is precisely in the sparse regime that we expect our approach to be useful for increasing the SNR, so that in the regime of short temporal windows, the experiment can be seen as a proof of principle, which validates the theoretical predictions of previous sections.

An overview of the four experimental configurations, together with the fit results analyzed below, is provided in Table~\ref{tab:fits}. \rmedit{The first measurement is performed with strong ambient illumination and the pump laser switched off (background noise only).} The second is performed with the pump operated at \SI{35}{mA} and the ambient illumination removed (signal plus dark counts only). The remaining two measurements are performed with the pump and ambient illumination present simultaneously, at pump currents of \SI{25.4}{mA} (weak signal) and \SI{35}{mA} (strong signal). For each configuration, the singles and coincidence rates are recorded across different postselection-window settings, without changing the optical alignment or detector settings. The acquisition software reports the singles rates $A$ and $B$ in the two detection channels and a quantity labeled A\&B, which we denote by $C_{AB}$. This quantity includes contributions from genuine photon pairs and accidental temporal associations between unrelated detections. We retain this software-reported rate throughout the experimental analysis.

\begin{figure}[t]
    \centering
    \includegraphics[width=\columnwidth]{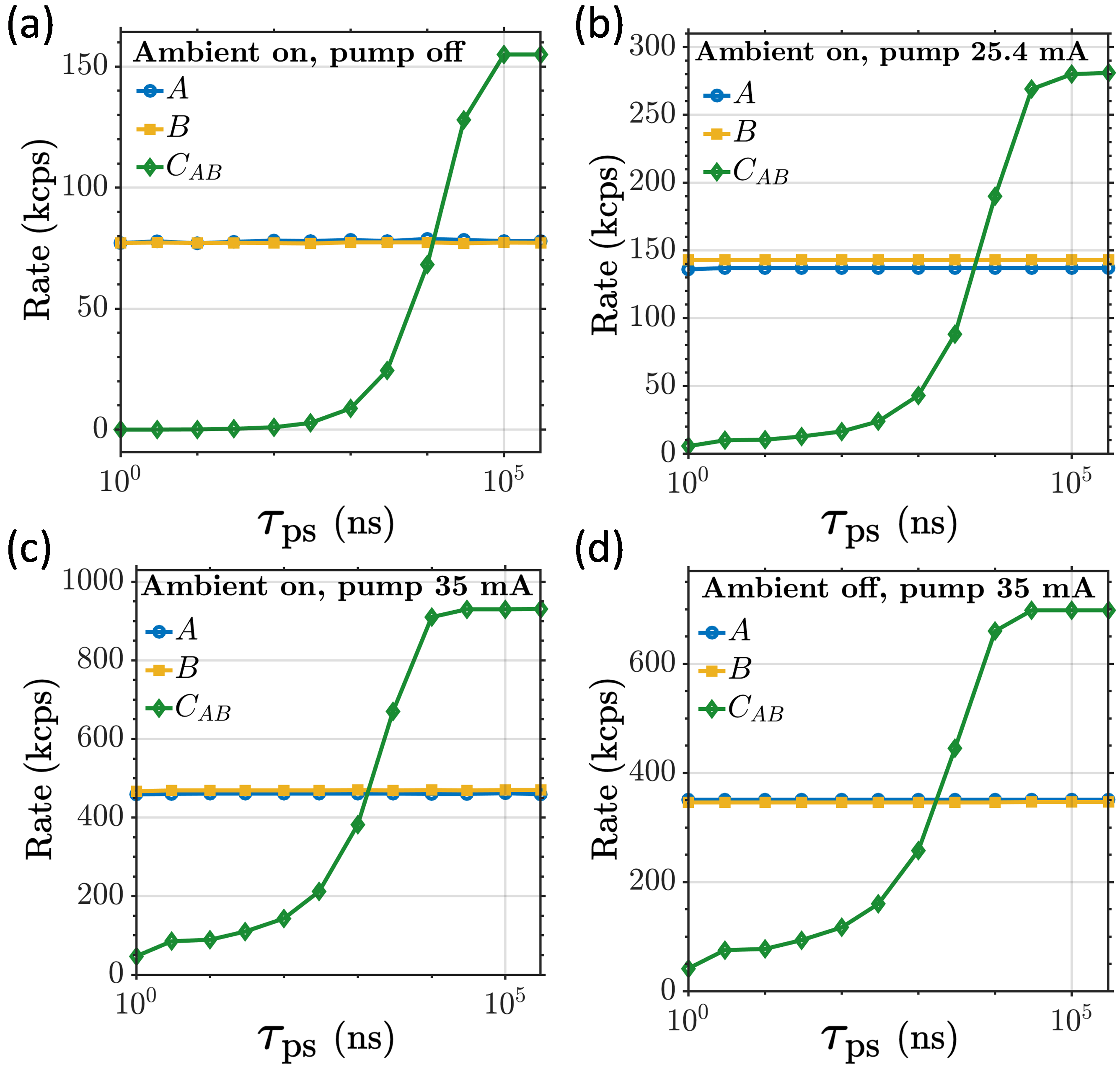}
    \caption{Measured singles rates \(A\) and \(B\), together with the retained-click rate \(C_{AB}\), as functions of the postselection window \(\tau_{\mathrm{ps}}\). The four panels correspond to (a) ambient illumination on with the pump off, (b) ambient illumination on with the pump operated at \(25.4\,\mathrm{mA}\), (c) ambient illumination on with the pump operated at \(35\,\mathrm{mA}\), and (d) ambient illumination off with the pump operated at \(35\,\mathrm{mA}\). }
    \label{fig:rates_vs_postselection}
\end{figure}
\begin{table*}[t]
\caption{Quadratic fits of the retained-click rate $C_{AB}=a+b\tps+c\tps^2$ over the interval from \SI{3}{ns} to \SI{300}{ns}. The rate is expressed in kcps and $\tps$ in ns. The coefficients provide a phenomenological description of the short-window data, while the RMSE quantifies the residual deviation from the fitted curve.}
\label{tab:fits}
\begin{ruledtabular}
\begin{tabular}{lccccc}
Configuration & $a$ (kcps) & $b$ (kcps/ns) & $c$ (kcps/ns$^2$) & $R^2$ & RMSE (kcps)\\
\hline
Ambient on, pump off
& 0.01866 & 0.01001 & $-2.792\times10^{-6}$ & 0.99981 & 0.0221\\
Ambient off, pump \SI{35}{mA}
& 74.713 & 0.51308 & $-7.639\times10^{-4}$ & 0.99479 & 3.583\\
Ambient on, pump \SI{25.4}{mA}
& 9.7989 & 0.07884 & $-1.074\times10^{-4}$ & 0.99586 & 0.523\\
Ambient on, pump \SI{35}{mA}
& 84.468 & 0.69327 & $-8.956\times10^{-4}$ & 0.99619 & 4.572\\
\end{tabular}
\end{ruledtabular}
\end{table*}

\subsection{Window dependence and saturation}
The results of table \ref{tab:fits} are plotted in Fig. \ref{fig:rates_vs_postselection} and~\ref{fig:normalized_and_absolute_rates}. The rates of individual detectors (single photons) remain approximately constant over the explored range of \(\tau_{\mathrm{ps}}\), whereas \(C_{AB}\) increases as a progressively larger fraction of temporally separated clicks is retained.  \rmedit{At short time windows, the ambient-only configuration is strongly suppressed}, while the pump-driven configurations retain a substantial contribution, reflecting the presence of temporally correlated photon-pair events.

For large $\tps$, the measured rate $C_{AB}$ approaches the limiting value $A+B$. In this limit, nearly all clicks are accepted by the software's poor-matching procedure. This artificial saturation is outside the sparse-window regime of our theoretical model; the short-window measurements are the appropriate basis for comparison. The contrast between ambient-only and pump-driven measurements is already apparent in the regime of short temporal windows: \rmedit{below \SI{100}{\nano\second} (left portions of Figs.~\ref{fig:rates_vs_postselection} and \ref{fig:normalized_and_absolute_rates})}. With the pump switched off, $C_{AB}$ is strongly suppressed at short windows despite singles rates of approximately \SI{77}{kcps} in each channel. Activating the pump produces a much larger retained-click contribution at the shortest measured windows, \rmedit{consistent with significant production of twin photons}. The persistence of this contribution under simultaneous pump and ambient illumination provides the experimental basis for detecting a weak source of simultaneous pairs hidden in a strong noisy background.

\subsection{\rmedit{Short time-window behavior}}
To quantify the pre-saturation behavior, in the sparse regime of interest, we fitted the five measurements between \SI{3}{ns} and \SI{300}{ns} with the phenomenological expansion
\begin{equation}
C_{AB}(\tps)=a+b\tps+c\tps^2.
\label{eq:polyfit}
\end{equation}
The coefficient $a$ captures the portion of the measured rate that varies weakly over the fitted interval and is therefore phenomenologically associated with the genuine pair production rate.  The coefficient $b$ contains contributions \rmedit{from all kinds} of accidental pairs. The quadratic term empirically accounts for the curvature away from the linear sparse-window behavior due to the aforementioned saturation effects. The measured short-window data and the corresponding quadratic fits are shown in Fig.~\ref{fig:short_window_fits}.

\rmedit{The fitted parameters are summarized in Table~\ref{tab:fits}. In its $a$ column, $a_N=\SI{0.01866}{kcps}$ corresponds to ambient illumination only (pump off), $a_L=\SI{74.713}{kcps}$ to the $\SI{35}{mA}$ pump only (ambient off), and $a_{L+N}=\SI{84.468}{kcps}$ to that pump with ambient illumination. The weak-pump ($\SI{25.4}{mA}$) row is a separate configuration and is not included in the $L+N$ comparison in the Appendix.} For ambient illumination alone, the estimate of the coincidence rate in the limit where the window width goes to zero is $a_N=\SI{0.01866}{kcps}$, whereas the pump-driven measurements give values between \SI{9.8}{kcps} and \SI{84.5}{kcps}. This separation is consistent with the short-window behavior visible in Fig.~\ref{fig:short_window_fits}a: the ambient-only signal is dominated by coincidences that increase with the postselection window.

A more direct test is obtained from the data measured with the laser turned off (subpanel (a) in Fig.\ref{fig:short_window_fits} and first line of table~\ref{tab:fits}). For short coincidence windows, we compare the measured rate $C_{AB}$ with the value calculated from
\begin{equation}
C_{AB}\simeq 2AB\tps.
\label{eq:Cind}
\end{equation}                                                                                            
For $\tau=\SI{3}{ns}$, we measured $A=\SI{77.8}{kcps}$ and $B=\SI{77.3}{kcps}$, yielding $C_{AB}^{\mathrm{pred}}=2AB\tps\simeq\SI{0.0361}{kcps}$. The corresponding measured rate is $\SI{0.037}{kcps}$. The relative difference is approximately $2.5\%$. 
This agreement supports our prediction that the measured rate results from accidental matching between statistically independent detections Eq.~\eqref{eq:Cind}.


\section{QKD with pairs of maximally entangled simultaneous photons.}
\label{sec:experiment}

\subsection{Experimental setup and measurement protocol}

\rmedit{To generate polarization-entangled photon pairs, we used the polarization-entanglement extension  \cite{ThorlabsQOP} of the device described in the previous section.}

\begin{figure*}[t]
\centering
\includegraphics[width=0.9\textwidth]{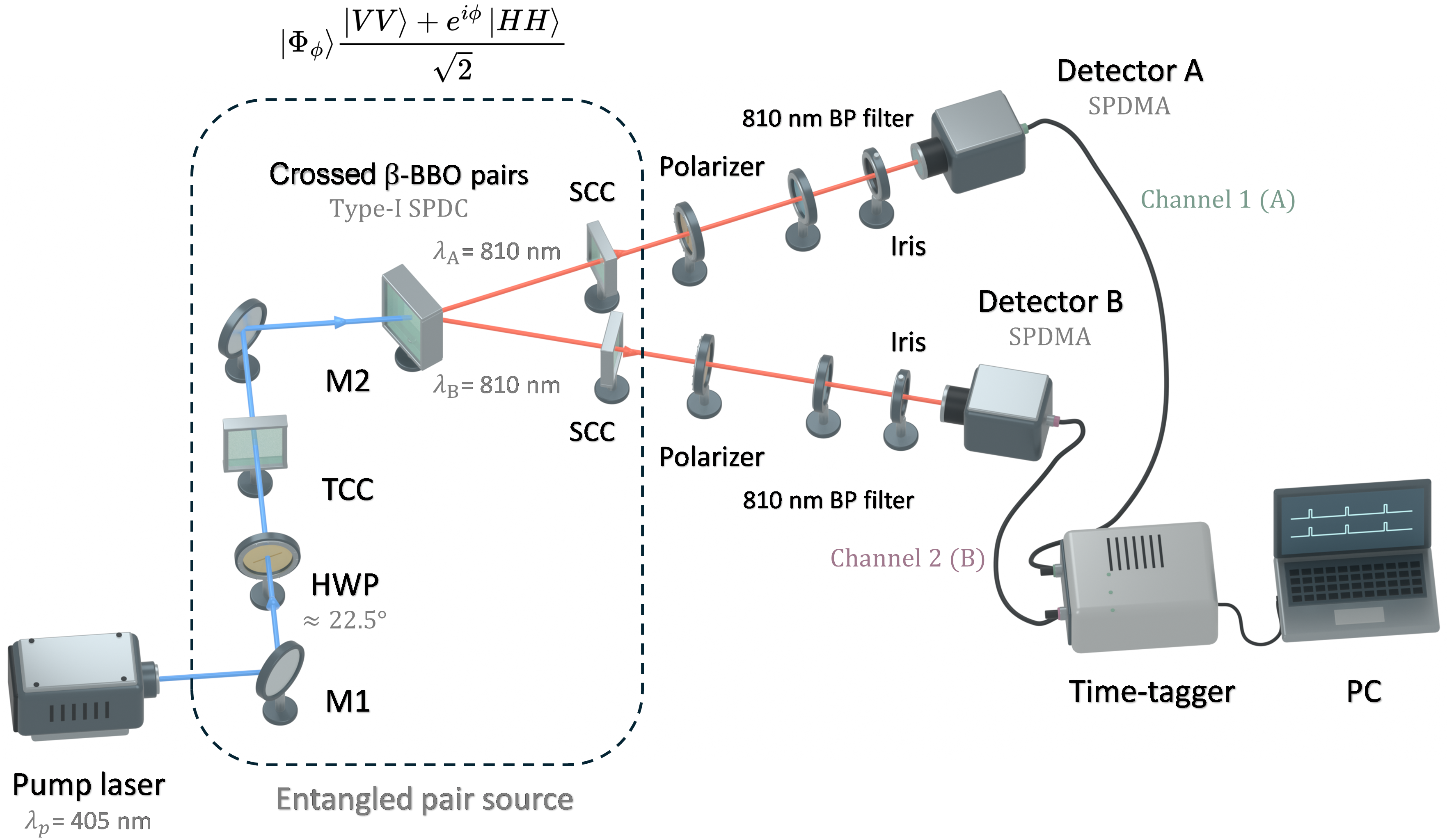}
\caption{Experimental setup for generating and measuring polarization-entangled photon pairs. A continuous-wave laser at \(405\,\mathrm{nm}\) is directed by mirrors M1 and M2 through a half-wave plate (HWP) and a temporal compensation crystal (TCC) onto two crossed type-I (\(\beta\)-BBO) crystals, producing photon pairs near \(810\,\mathrm{nm}\). The HWP sets the pump polarization, and the TCC and spatial compensation crystals (SCC) reduce temporal and spatial distinguishability between the two pair-generation processes. Each photon passes through a polarizer, an \(810\,\mathrm{nm}\) band-pass filter, and an iris before reaching a single-photon counting module. The detector pulses are recorded on separate channels of a time tagger, with the acquisition data transferred to a computer for polarization-correlation and temporal postselection analysis. }
\label{fig:experimental_setup}
\end{figure*}
The experimental arrangement is shown in Fig.~\ref{fig:experimental_setup}. A continuous-wave laser at \(405\,\mathrm{nm}\) is directed by mirrors M1 and M2 through a half-wave plate (HWP) and a temporal compensation crystal (TCC) onto two crossed type-I \(\beta\)-barium borate (\(\beta\)-BBO) crystals. The polarization of the pump is adjusted with a half-wave plate oriented close to \(22.5^\circ\), so that the two crystals are excited coherently and generate the two-photon state :
\begin{equation}
\ket{\Phi_\phi}=\frac{\ket{VV}+e^{i\phi}\ket{HH}}{\sqrt{2}},
\end{equation}
\rmedit{where the phase \(\phi\) depends on the orientation of the compensation crystals, TCC and SCC \cite{Kwiat1999}.} 

The twin photons are now correlated in polarization. We adjusted the phase to $\phi\approx0$ to approach the target state $\ket{\Phi_0}=(\ket{VV}+\ket{HH})/\sqrt{2}$. The detection arrangement selects the measurement basis and the optical channels through which the photon state is probed \cite{hatifi2026e}. We distinguish two analyzer configurations. Measurements with both analyzers set to $V$ probe the correlated $VV$ outcome and the recovery of its intrinsic pair contribution relative to the ambient background. Separate measurements with analyzer $A$ set to $H$ and analyzer $B$ set to $V$ probe the $HV$ outcome.

We model state-preparation imperfections by the Werner state of Eq.~\eqref{eq:Werner}, with $\ket{\Psi}=\ket{\Phi_0}$. The corresponding intrinsic probabilities are $p_{VV}=p_{HH}=1/2-\epsilon/4$ and $p_{HV}=p_{VH}=\epsilon/4$. The separately added ambient illumination produces accidental coincidences and does not, by itself, change the state parameter $\epsilon$. Thus, the intrinsic $VV$ coincidence rate is a correlated-outcome rate, whereas the intrinsic $HV$ rate probes an error outcome.

\subsection{\texorpdfstring{Correlated-outcome measurements in the $VV$ configuration}{Correlated-outcome measurements in the VV configuration}}

We performed measurements in the $VV$ configuration, with both polarization analyzers set to $V$. Finite timing resolution prevents direct access to the zero-window limit. We therefore fit the measured $C_{VV}(\tps)$ in the sparse-window range using $C_{VV}(\tps)=C_{VV}^{\mathrm{int}}+b\tps$ and extrapolate the fitted lines to $\tps=0$. This intercept estimates the window-independent contribution, under the assumption that genuine pair detections are retained over the fitted range, while accidental associations scale approximately linearly with $\tps$.

\begin{figure}[!t]
\centering
\includegraphics[width=\linewidth]{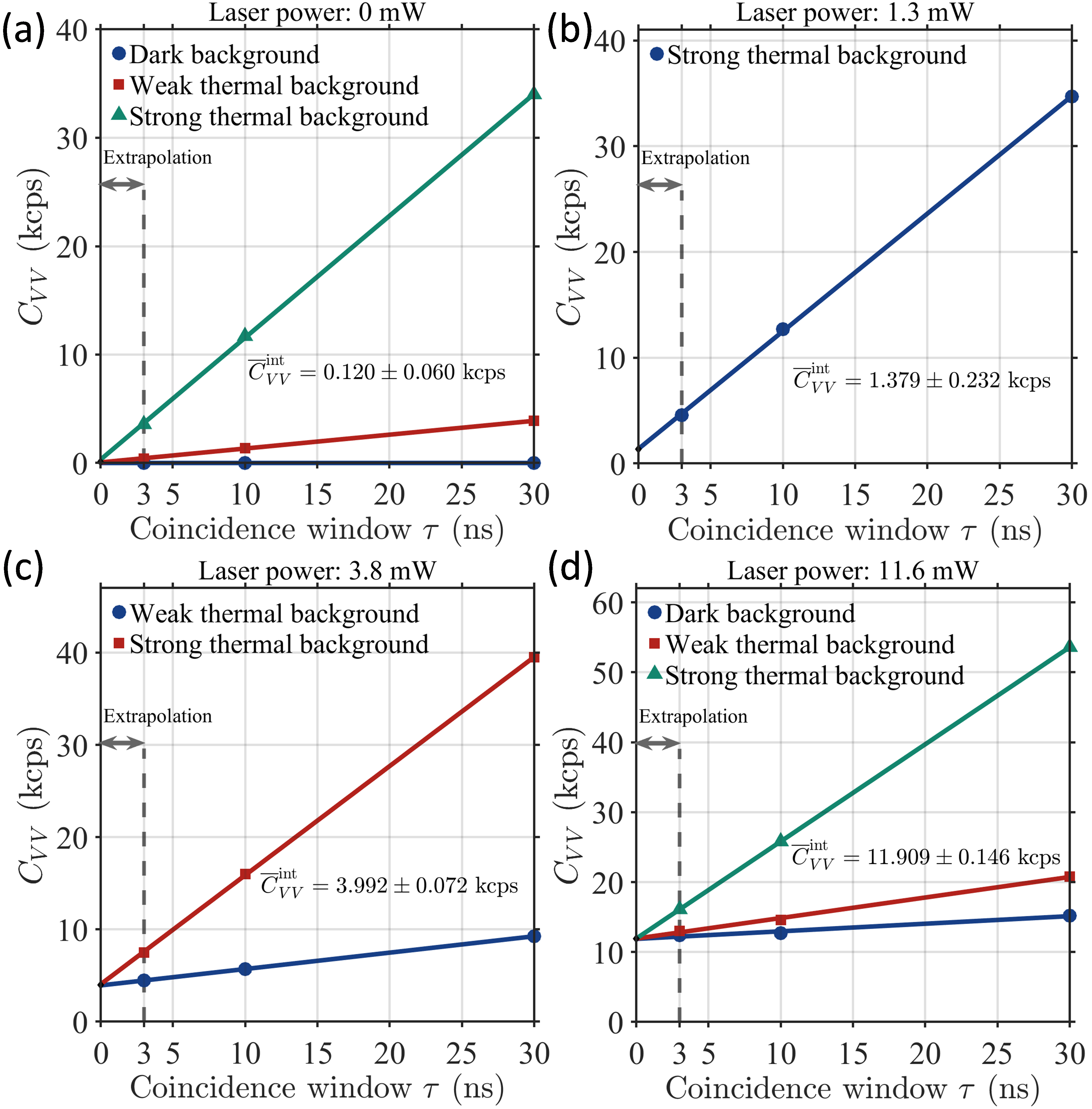}
\caption{Measured $VV$ rates $C_{VV}$ versus postselection-window width $\tau$, with both polarization analyzers set to $V$, at nominal pump powers of (a) 0, (b) 1.3, (c) 3.8, and (d) 11.6 mW. Symbols show measurements at $\tau=3$, 10, and 30 ns under the background conditions indicated in each legend. Solid lines are separate linear fits to these three points, extrapolated to $\tau=0$. Dashed vertical lines mark the shortest fitted window, 3 ns, and grey arrows indicate the extrapolation interval. The annotations report the mean fitted intercept $\overline{C}_{VV}^{\mathrm{int}}$ across the background conditions at each power; panel (b) contains a single fit. Quoted uncertainties are formal fit-derived standard errors, propagated when averaging intercepts.}
\label{fig:VV_population_pump_power}
\end{figure}

\begin{figure}[!t]
\centering
\includegraphics[width=\linewidth]{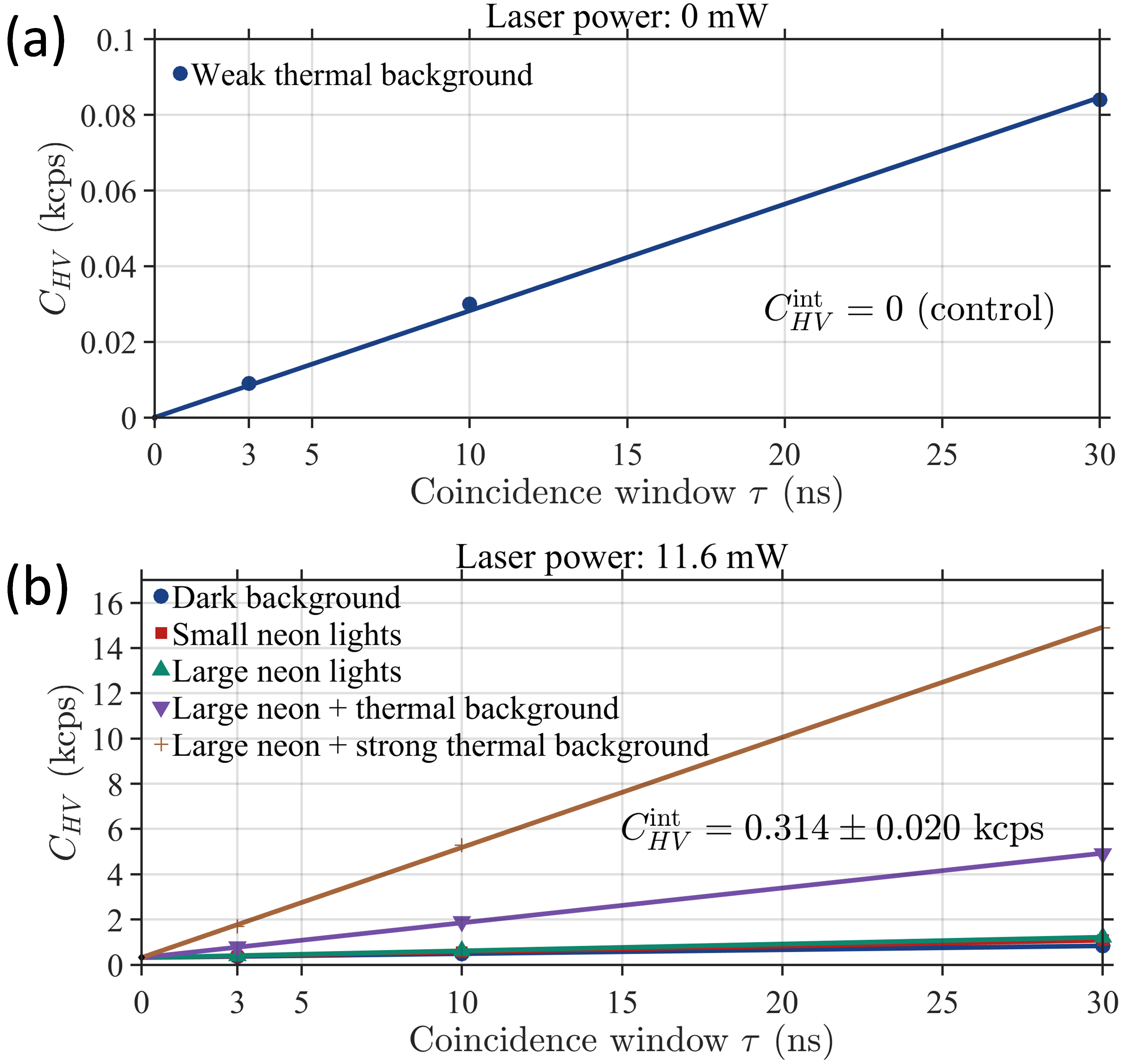}
\caption{Measured $HV$ rates $C_{HV}$ versus postselection-window width $\tau$, with the polarization analyzers set to $H$ and $V$. Symbols show measurements at $\tau=3$, 10, and 30 ns; solid lines are linear least-squares fits extrapolated to $\tau=0$. (a) Pump-off control under weak thermal background, with the fitted intercept constrained to zero. (b) Measurements at a nominal pump power of 11.6 mW under the five background conditions indicated in the legend. A joint fit with a common intercept and a separate slope for each background gives $C_{HV}^{\mathrm{int}}=0.314\pm0.020$ kcps. The common intercept is imposed by the fitting model, and the quoted uncertainty is its formal fit-derived standard error.}
\label{fig:HV_window_dependence}
\end{figure}


The four series in Fig.~\ref{fig:VV_population_pump_power} correspond to different nominal pump powers. With the pump off Fig.~\ref{fig:VV_population_pump_power}(a), we measured a dark-room control and two additional controls with weak and strong thermal illumination. At \SI{11.6}{mW}, Fig.~\ref{fig:VV_population_pump_power}(b), measurements were performed in a dark room and with weak and strong thermal illumination. The fitted intercepts are close to one another, with a mean of approximately \SI{11.91}{kcps}. At \SI{3.8}{mW}, Fig.~\ref{fig:VV_population_pump_power}(c), the two runs used weak and strong thermal illumination. Their mean fitted intercept is approximately \SI{3.99}{kcps}. At \SI{1.3}{mW}, Fig.~\ref{fig:VV_population_pump_power}(d), only one run was recorded, with strong thermal illumination. Its fitted intercept is approximately \SI{1.38}{kcps}. For each background condition, a separate ordinary least-squares fit was performed using the three points at $\tps\in\{\SI{3}{ns},\SI{10}{ns},\SI{30}{ns}\}$. When several runs were available at a given pump power, the quantity $\overline{C}_{VV}^{\,\mathrm{int}}$ reported in Table~\ref{tab:theory_experiment} is the arithmetic mean of their fitted intercepts; for \SI{1.3}{mW}, it is the intercept of the single fit. The fits were not constrained to share a common intercept.

Standard errors of the individual intercepts were estimated from the residual covariance of the three-point unweighted fits. For independent runs, the standard error of their arithmetic mean was calculated as
\begin{equation}
\sigma_{\overline{C}}=\frac{\sqrt{\sum_{j=1}^{n}\sigma_j^2}}{n}.
\end{equation} The resulting $1\sigma$ regression uncertainties are approximately $0.060$, $0.146$, $0.072$, and $0.232$ kcps at $0$, $11.6$, $3.8$, and $1.3$ mW, respectively.

The normalized intrinsic $VV$ rates approximately follow the relative pump powers, as shown in Table~\ref{tab:theory_experiment}. This supports the recovery of a pump-dependent correlated-pair contribution despite varying ambient background. Combined with the separate $HV$ measurements, they allow an estimate of the intrinsic error under the symmetry assumptions described below.

\begin{table}[t]
\caption{Comparison between the relative nominal pump powers and the relative
experimentally inferred intrinsic coincidence rates,
$\overline{C}_{VV}^{\,\mathrm{int}}$.}
\label{tab:theory_experiment}
\centering
\renewcommand{\arraystretch}{2.1}
\setlength{\tabcolsep}{5pt}

\resizebox{\columnwidth}{!}{%
\begin{ruledtabular}
\begin{tabular}{ccc}
\shortstack{Pump power\\$P$ (mW)} &
\shortstack{Normalized pump power\\$P/P_{\max}$} &
\shortstack{Normalized intrinsic\\coincidence rate\\
$\overline{C}_{VV}^{\,\mathrm{int}}/
 \overline{C}_{VV,\max}^{\,\mathrm{int}}$}
\\
\colrule
$0.0$, Fig.~\ref{fig:VV_population_pump_power}(a) &
$\dfrac{0.00}{11.60}=0.00$ &
$\dfrac{0.12}{11.91}\approx 0.01$
\\
$1.3$, Fig.~\ref{fig:VV_population_pump_power}(d) &
$\dfrac{1.30}{11.60}\approx 0.11$ &
$\dfrac{1.38}{11.91}\approx 0.12$
\\
$3.8$, Fig.~\ref{fig:VV_population_pump_power}(c) &
$\dfrac{3.80}{11.60}\approx 0.33$ &
$\dfrac{3.99}{11.91}\approx 0.34$
\\
$11.6$, Fig.~\ref{fig:VV_population_pump_power}(b) &
$\dfrac{11.60}{11.60}=1.00$ &
$\dfrac{11.91}{11.91}=1.00$
\end{tabular}
\end{ruledtabular}%
}
\end{table}

\subsection{\texorpdfstring{Error-outcome measurements in the $HV$ configuration}{Error-outcome measurements in the HV configuration}}

A separate measurement series used $H$ in channel $A$ and $V$ in channel $B$. In this series, all recorded pump-on runs used the same nominal power of \SI{11.6}{mW}, while the ambient illumination was varied: dark room, small neon lights, large neon lights, and large neon lights supplemented by thermal illumination. The pump-off control used weak thermal illumination.

The pump-off control is shown in Fig.~\ref{fig:HV_window_dependence}(a). With the pump switched off and weak thermal illumination present, the measured $HV$ rate increases approximately linearly with the postselection window. The control fit is constrained to pass through zero.

Figure~\ref{fig:HV_window_dependence}(b) shows the pump-on $HV$ measurements at the fixed nominal power of \SI{11.6}{mW}. Different ambient-background conditions produce different slopes, whereas the plotted lines share the intrinsic intercept reported by the analysis, $\overline{C}_{HV}^{\,\mathrm{int}}=\SI{0.314 \pm 0.020}{kcps}$. The short-window dependence is represented by
\begin{equation}
C_{HV,j}^{\mathrm{fit}}(\tps)=\overline{C}_{HV}^{\,\mathrm{int}}+b_j\tps,
\label{eq:HV_window_fit}
\end{equation}
where $j$ labels the ambient-background condition.

Let $f_{ij}^{0}$ denote the intrinsic coincidence-event rate for outcome $ij$, inferred from short-window extrapolation. If all four rates are obtained under matched source conditions and the relative detection efficiencies are accounted for, the Werner-state model gives
\begin{equation}
\frac{\epsilon}{4}=\frac{f_{HV}^{0}}{f_{HH}^{0}+f_{HV}^{0}+f_{VH}^{0}+f_{VV}^{0}}.
\label{eq:epsilon_from_HV}
\end{equation}
To estimate $\epsilon$ from the available $VV$ and $HV$ measurements, we assume symmetry between the two correlated outcomes, $f_{HH}^{0}\simeq f_{VV}^{0}$, and between the two error outcomes, $f_{VH}^{0}\simeq f_{HV}^{0}$. We also assume comparable source brightness and effective detection efficiencies for the two measurement series at the same nominal pump power. Under these assumptions, Eq.~\eqref{eq:epsilon_from_HV} reduces to
\begin{equation}
\frac{\epsilon}{4}\simeq
\frac{\overline{C}_{HV}^{\,\mathrm{int}}}
{2\overline{C}_{VV}^{\,\mathrm{int}}+2\overline{C}_{HV}^{\,\mathrm{int}}}.
\label{eq:epsilon_from_VV_HV_symmetry}
\end{equation}
At a nominal pump power of \SI{11.6}{mW}, the $VV$ fits give $\overline{C}_{VV}^{\,\mathrm{int}}\simeq\SI{11.91}{kcps}$, while the $HV$ analysis in Fig.~\ref{fig:HV_window_dependence}(b) gives $\overline{C}_{HV}^{\,\mathrm{int}}\simeq\SI{0.314}{kcps}$. Consequently,
\begin{equation}
\epsilon\simeq
4\times\frac{0.314}{11.91+0.314+0.314+11.91}
\simeq0.051.
\end{equation}
The estimated intrinsic probability of a single $HV$ outcome is therefore approximately $1.3\%$. The intrinsic QBER includes both error outcomes, $HV$ and $VH$, and is
\begin{equation}
\QBER_{\mathrm{int}}\simeq
\frac{f_{HV}^{0}+f_{VH}^{0}}{\sum_{i,j\in\{H,V\}}f_{ij}^{0}}
=\frac{\epsilon}{2}\simeq2.6\%.
\label{eq:QBER_experimental_symmetry}
\end{equation}


\FloatBarrier
\section{Conclusions.}
\label{sec:conclusions}
To conclude, experimental data confirmed our theoretical analysis: reducing the width of the postselection window increases the signal-to-noise ratio because the intrinsic signal is insensitive to this postselection, whereas the probability of accidental pair detections scales proportionally to the window width. For entanglement-based quantum key distribution, for instance, the extrapolated rates yield an estimated intrinsic QBER of approximately $2.6\%$. This information can guide strategies for improving retained-bit quality. \rmedit{Our analysis estimates the contributions of different noise sources and the potential benefit of improving experimentally controllable parameters.}

Suppose that the intrinsic QBER in the Earth--Moon example of Sec.~\ref{sec:moonmain} is $8\%$. From Eqs.~\eqref{eq:moontrue} and \eqref{eq:moonaccidentals}, $\Rtrue\simeq\SI{10}{s^{-1}}$ and $\Racc\simeq\SI{1}{s^{-1}}$. An accidental coincidence gives a random bit, which is wrong with probability $1/2$; its contribution to the total QBER for an otherwise ideal state is therefore $(0.5\Racc)/(\Rtrue+\Racc)\simeq4.5\%$. Including the $8\%$ error, we get approximately $12\%$. The result exceeds the $11\%$ upper bound mentioned above  \cite{Cerf2002}. The security of the entanglement-based protocol requires an analysis appropriate to it. If the detection system could resolve $\tps=\SI{1e-13}{s}$ while retaining genuine pairs, Eq.~\eqref{eq:Rmp} predicts a tenfold reduction of the cross-pair rate; under unchanged source and channel parameters, the same simplified model gives a total QBER of approximately $8.4\%$, still within the safety threshold.

Regarding anti-counterfeiting, our ideas are also useful for experimentally revealing the presence of a very weak source of simultaneously emitted photon pairs. Even if, due to the presence of strong ambient noise and various sources of accidental simultaneous clicks (dark count, fluorescence, and so on), the genuine signal remains hidden within a noisy background, our method makes it possible to estimate the intrinsic pair-production rate. Our extrapolation method thus makes it possible to artificially eliminate the aforementioned sources of noise that contaminate our measurements.

Of course, the experimental results described in Sec.~\ref{sec:experiment1} constitute only a proof of principle. The present experiment relies on a bright bulk-crystal SPDC source \cite{Kwiat1999,ThorlabsQOP}, whereas practical anticounterfeiting and tagging schemes will likely require compact and integrable quantum-light platforms. Candidate approaches investigated within the ARTEMIS consortium include molecular lanthanide emitters produced by the University of Eastern Piedmont team in Vercelli \cite{Bonabello2026}, nonlinear molecular materials developed in Cagliari \cite{Attar2025}, and plasmonically engineered nanostructures explored by the Vigo team \cite{Santiago2025}. Other promising platforms include ultrathin nonlinear metasurfaces \cite{Zhang2022}, resonant metasurfaces \cite{SantiagoCruz2021} and semiconductor quantum dots \cite{Orieux2017}.
To upgrade our technique to the prototype level requires to characterize and to identify sufficiently \rmedit{bright yet hidden and discrete sources} of simultaneous photon pairs (or even triplets or quadruplets, as explained in section \ref{sec:multiplets}). Obviously, the crystals used in our experiment are not easy to hide, in contrast to nanocomponents, which are smaller to begin with. Phase-matching -see Fig. \ref{fig:scheme_1}  is more difficult to realize in this case, and we expect that fewer pairs will be produced and that they will be scattered over a large range of directions. Actually, second harmonic generation has been observed in many nonlinear nanocomponents \cite{Bachelier2010}, but its inverse process, parametric down-conversion, is much less documented due to its intrinsic weakness \cite{SantiagoCruz2021}. Therefore, we also expect the signal to be hidden in a consequent background noise. As a first step, the extrapolation method of Sec.~\ref{sec:experiment1} will be used to estimate the intrinsic pair rate while suppressing window-dependent accidental contributions. From this perspective, our approach can be considered a cornerstone of anticounterfeiting and marking based on paired sources.

 \section*{Acknowledgments}
 
We thank Benoit Chalopin for helpful discussions. Ce travail a bénéficié d'une aide de l’État gérée par l'Agence Nationale de la Recherche au titre de France 2030 et par l’Union européenne NextGenerationEU portant la référence ANR-22-CMAS-0001 (QuanTEdu-France). This work was carried out within the framework of the ARTEMIS project, ``Molecular Materials for On-Chip Integrated Quantum Light Sources,'' and was supported by the European Union through the EIC Pathfinder Challenges 2022 programme under Grant Agreement No.~101115149. The content reflects only the authors' views, and the European Commission is not responsible for any use that may be made of the information it contains.

%

\AtEndEnvironment{thebibliography}{%
  \bibitem[Hatifi(2026)]{Hatifi2026Geometry}%
  M.~Hatifi,
  \href{https://doi.org/10.1103/mycc-r3bx}{Geometry-controlled freezing and revival of Bell nonlocality through environmental memory},
  {\it Phys. Rev. A} {\bf 113}, 022204 (2026).
}

\clearpage
\onecolumngrid

\appendix
\section{Mixed accidental pairs.}
\begin{figure}[!htbp]
    \centering
    \includegraphics[width=0.60\linewidth]{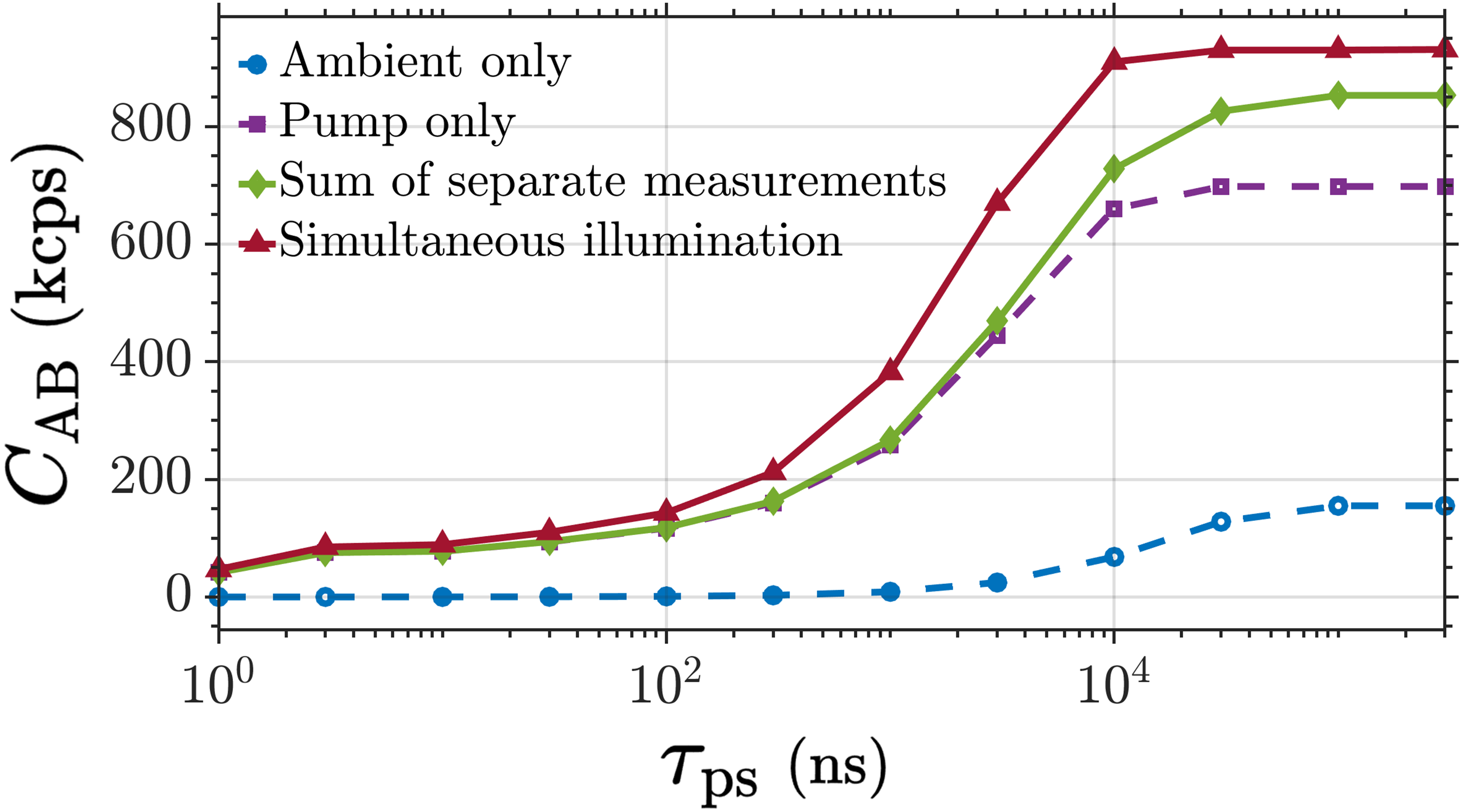}
    \caption{Test of additivity for the correlation rate $C_{AB}$. The dashed blue and purple curves show the measurements obtained with ambient illumination alone and with the $35\,\mathrm{mA}$ pump alone, respectively. The green curve is the pointwise sum of these two separate measurements, while the red curve corresponds to the simultaneous presence of ambient illumination and the pump. The red curve systematically exceeds the green curve.}
    \label{fig:additivity_test}
\end{figure}

The results plotted in Fig.~\ref{fig:additivity_test} also provide another direct test of our theoretical model of Sec.~\ref{sec:marking}. They clearly establish that the rates of simultaneous events collected with the pump and the external source of ambient light turned on simultaneously is higher than the sum of their separate contributions, demonstrating the presence of accidental pairs involving one ambient-induced click and one pump-induced click. As always, at sufficiently large \(\tau_{\mathrm{ps}}\), all curves approach their respective saturation values as the matching procedure retains nearly all detected clicks.

We denote the ambient-only singles rates by $A_N$ and $B_N$, the pump-only rates by $A_L$ and $B_L$, and the rates measured under simultaneous illumination by $A_{L+N}$ and $B_{L+N}$. Because the independent retained-click slope is proportional to the product of the singles rates, the excess slope expected in the simultaneous acquisition is (\ref{eq:SNRmarkerExplicit})

\begin{equation}
b_{\times}^{(\mathrm{pred})}
=2\times10^{-6}\left(
\overline A_{L+N}\overline B_{L+N}
-\overline A_L\overline B_L
-\overline A_N\overline B_N
\right),
\label{eq:bcrosspred}
\end{equation}
where the numerical prefactor converts singles rates in kcps and window widths in ns into a slope in kcps/ns. When the simultaneous singles equal the sums of the separately measured pump and ambient contributions, this expression reduces to the two mixed products proportional to $A_LB_N+A_NB_L$.

Averaging the singles rates over the fitted interval gives
\begin{align}
(\overline A_N,\overline B_N)&=(77.68,77.12)\,\mathrm{kcps},\\
(\overline A_L,\overline B_L)&=(351.0,346.0)\,\mathrm{kcps},\\
(\overline A_{L+N},\overline B_{L+N})&=(460.8,469.0)\,\mathrm{kcps}.
\end{align}
Equation~\eqref{eq:bcrosspred} then predicts
\begin{equation}
b_{\times}^{(\mathrm{pred})}=0.1774\,\mathrm{kcps/ns}.
\label{eq:bcrosspredvalue}
\end{equation}
The same quantity extracted from the fitted slopes is
\begin{equation}
b_{\times}^{(\mathrm{fit})}
=b_{L+N}-b_L-b_N
=0.1702\,\mathrm{kcps/ns},
\label{eq:bcrossfit}
\end{equation}
which differs from the singles-product prediction by approximately $4.1\%$. The corresponding nonadditivity over the full postselection-window range is shown in Fig.~\ref{fig:additivity_test}. The agreement provides a quantitative consistency check that the excess window-dependent contribution under simultaneous illumination is governed predominantly by mixed associations between pump-induced and ambient detections. 

This excess has different operational meanings in the two applications considered here. In QKD, a coincidence formed by one surviving pair of photons and one background click carries a random bit value and contributes to the error rate. In the marker measurement, the same term is absent when the pump-driven source is removed, thereby contributing to the marker-dependent on-off signal. 

\clearpage
\section{Detailed description of the data analysis}
\label{app:data-analysis}

\begin{figure}[!htbp]
    \centering
    \includegraphics[width=\linewidth]{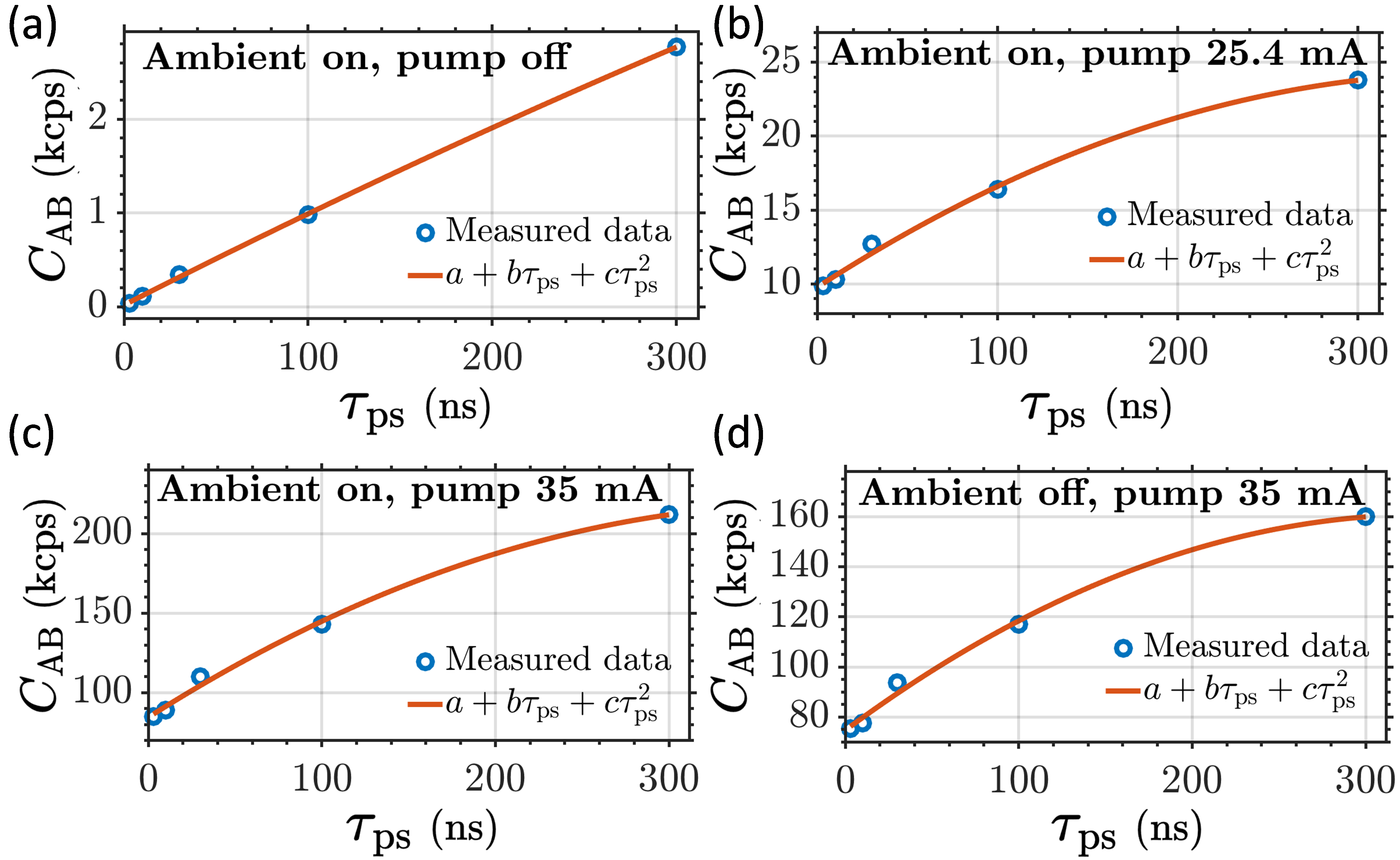}
    \caption{Pair detection rate $C_{AB}$ in the sparse regime for the four experimental configurations. The symbols show the measured data, while the solid curves are phenomenological quadratic fits of the form $C_{AB}=a+b\tau_{\mathrm{ps}}+c\tau_{\mathrm{ps}}^{2}$ over the interval $\SI{3}{ns}\leq\tau_{\mathrm{ps}}\leq\SI{300}{ns}$.}
    \label{fig:short_window_fits}
\end{figure}

\begin{figure}[!htbp]
    \centering
    \includegraphics[width=0.75\linewidth]{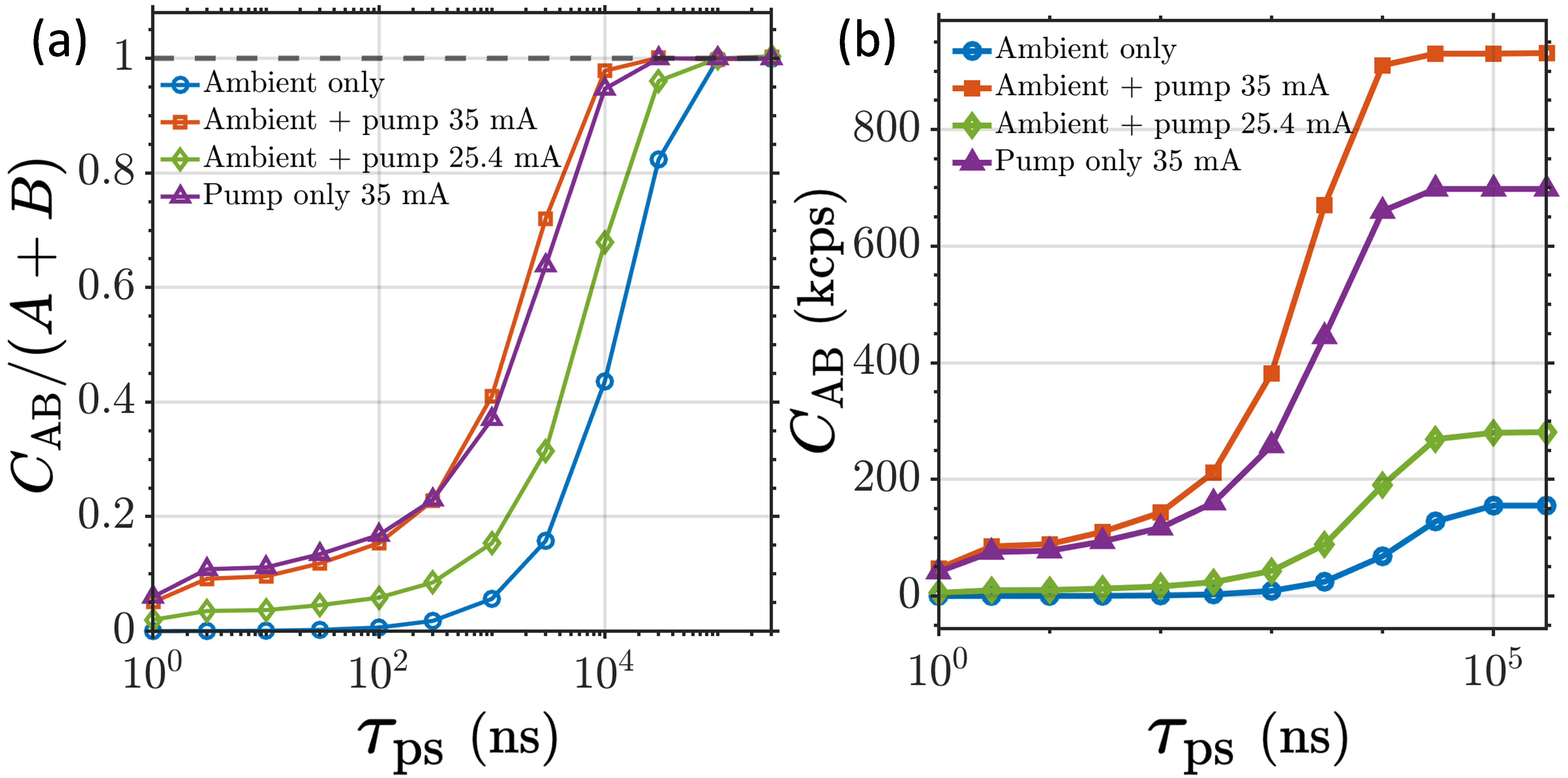}
    \caption{Postselection-window dependence of the retained-click signal for the four experimental configurations. (a) Normalized fraction $C_{AB}/(A+B)$, where $A$ and $B$ are the measured singles rates in the two detection channels. The dashed horizontal line marks the saturation limit $C_{AB}=A+B$, reached when the postselection window becomes sufficiently large that essentially every detected click can be associated with a click in the opposite channel. (b) Same as the left figure but this time with absolute, not normalized pair rates $C_{AB}$ for the same four configurations.}
    \label{fig:normalized_and_absolute_rates}
\end{figure}

\clearpage

\end{document}